\documentclass[italian,english]{article}
\usepackage[latin9]{inputenc}
\usepackage{color}
\usepackage{graphicx}
\usepackage{amssymb}

\newcommand{\lyxaddress}[1]{
\par {\raggedright #1
\vspace{1.4em}
\noindent\par}
}

\usepackage{babel}

\begin{document}

\title{\textbf{A precise response function for the }\textbf{\emph{magnetic}}\textbf{
component of Gravitational Waves in Scalar-Tensor Gravity}}

\author{\textbf{Christian Corda}}

\maketitle

\lyxaddress{\begin{center}
International Institute for Theoretical Physics and Mathematics Einstein-Galilei,
Via Santa Gonda, 14 - 59100 PRATO, Italy %
\footnote{\begin{quotation}
\emph{Partially supported by a Research Grant of the R. M. Santilli
Foundations Number RMS-TH-5735A2310}
\end{quotation}
}
\par\end{center}}

\lyxaddress{\begin{center}
\textit{E-mail addresses:} \textcolor{blue}{cordac.galilei@gmail.com} 
\par\end{center}}
\begin{abstract}
The important issue of the \emph{magnetic} component of gravitational
waves (GWs) has been considered in various papers in the literature.
From such analyses, it resulted that such a \emph{magnetic} component
becomes particularly important in the high frequency portion of the
frequency range of ground based interferometers for GWs which arises
from standard General Theory of Relativity (GTR).

Recently, such a \emph{magnetic} component has been extended to GWs
arising from Scalar-Tensor Gravity (STG) too. After a review of some
important issues on GWs in STG, in this paper we re-analyse the \emph{magnetic}
component in the framework of STG from a different point of view,
by correcting an error in a previous paper and by releasing a more
precise response function. In this way, we also show that if one neglects
the \emph{magnetic} contribution considering only the low-frequency
approximation of the \emph{electric} contribution, an important part
of the signal could be, in principle, lost. The determination of a
more precise response function for the \emph{magnetic} contribution
is important also in the framework of the possibility to distinguish
other gravitational theories from GTR. 

At the end of the paper an expansion of the main results is also shown
in order to recall the presence of the magnetic component in GRT too.
\end{abstract}

\lyxaddress{\textit{Keywords}: scalar gravitational waves; interferometers; magnetic
components.}

\lyxaddress{PACS numbers: 04.80.Nn, 04.80.-y, 04.25.Nx}

\section{Introduction}

The data analysis of interferometric GWs detectors has nowadays been
started, and the scientific community hopes in a first direct detection
of GWs in next years; for the current status of GWs interferometers
see Ref. \cite{key-1}. In such a way, the indirect evidence of the
existence of GWs by Hulse and Taylor \cite{key-2}, Nobel Prize winners,
will be confirmed. Detectors for GWs will be important for a better
knowledge of the Universe and also because the interferometric GWs
detection will be the definitive test for GTR or, alternatively, a
strong endorsement for Extended Theories of Gravity (ETG) \cite{key-3}.
On the other hand, the discovery of GW emission by the compact binary
system composed by two Neutron Stars PSR1913+16 \cite{key-2} has
been, for physicists working in this field, the ultimate thrust allowing
to reach the extremely sophisticated technology needed for investigating
in this field of research \cite{key-1}. GWs are a consequence of
Einstein's GTR \cite{key-4}, which presuppose GWs to be ripples in
the space-time curvature travelling at light speed \cite{key-5,key-6}.
In GTR only asymmetric astrophysics sources can emit GWs \cite{key-7}.
The most efficient are coalescing binaries systems at frequencies
around 1 KHz \cite{key-1}, while a single rotating pulsar can rely
only on spherical asymmetries, usually very small \cite{key-1,key-7}.
Its spin frequency often lie in the hectohertz \emph{{}``sweet spot}''
of current detectors, i.e. at order hundreds Hz \cite{key-49}. Supernovae
could have relevant asymmetries, being potential sources \cite{key-7}.
It is generally agreed that the frequency of GW emission from the
birth of stellar mass collapsed objects is in the range 50Hz to a
few KHz \cite{key-50}. The most important cosmological source of
GWs is, in principle, the so-called stochastic background of GWs which,
together with the Cosmic Background Radiation (CBR), would carry,
if detected, a huge amount of information on the early stages of the
Universe evolution \cite{key-8,key-9,key-10,key-11}. The existence
of a relic stochastic background of GWs is a consequence of generals
assumptions. Essentially it derives from a mixing between basic principles
of classical theories of gravity and of quantum field theory \cite{key-8,key-9,key-10,key-11}.
The strong variations of the gravitational field in the early universe
amplify the zero-point quantum oscillations and produce relic GWs.
It is well known that the detection of relic GWs is the only way to
learn about the evolution of the very early universe, up to the bounds
of the Planck epoch and the initial singularity \cite{key-8,key-11}.
It is very important to stress the unavoidable and fundamental character
of this mechanism. The model derives from the inflationary scenario
for the early universe \cite{key-12}, which is consistent with the
WMAP data on the CBR (in particular exponential inflation and spectral
index \ensuremath{\approx} 1 \cite{key-13}). Inflationary models
are cosmological models in which the Universe undergoes a brief phase
of a very rapid expansion in early times \cite{key-12}. In this tapestry
the expansion could be power-law or exponential in time. Such models
provide solutions to the horizon and flatness problems and contain
a mechanism which creates perturbations in all fields \cite{key-12}.
Important for our case is that this mechanism also provides a distinctive
spectrum of relic GWs \cite{key-8,key-10,key-11}. The GWs perturbations
arise from the uncertainty principle and the spectrum of relic GWs
is generated from the adiabatically-amplified zero-point fluctuations
\cite{key-8,key-9,key-10,key-11}. In standard cosmology such a spectrum
is flat along the frequency range $10^{-16}\leq f\leq10^{8}\mbox{ }Hz$
\cite{key-51}.

Regarding the potential GW detection, let us recall some historical
notes. In 1957, F. A. E. Pirani, who was a member of the Bondi's research
group, proposed the geodesic deviation equation as a tool for designing
a practical GW detector {[}14{]}. In 1959, Joseph Weber {[}15{]},
\foreignlanguage{italian}{First Award Winner at the 1959 Gravity Research
Foundation Competition,} studied a detector that, in principle, might
be able to measure displacements smaller than the size of the nucleus.
He developed an experiment using a large suspended bar of aluminium,
with a high resonant Q at a frequency of about 1 kHz. Then, in 1960,
he tried to test the general relativistic prediction of GWs from strong
gravity collisions {[}16{]} and, in 1969, he claimed evidence for
observation of GWs (based on coincident signals) from two bars separated
by 1000 km {[}17{]}. He also proposed the idea of doing an experiment
to detect GWs by using laser interferometers {[}17{]}. In fact, all
the modern detectors can be considered like being originated from
early Weber's ideas \cite{key-1,key-7,key-18}. At the present time,
in the world there are five cryogenic bar detectors which have been
built to work at very low temperatures ($<4K$): Explorer at CERN,
Nautilus at Frascati INFN National Laboratory, Auriga at Legnaro National
Laboratory, Allegro at Luisiana State University and Niobe in Perth
\cite{key-7,key-18}. Instrumental details can be found in \cite{key-18}
and references within. Spherical detectors are the Mario Schenberg,
which has been built in San Paolo (Brazil) and the MiniGRAIL, which
has been built at the Kamerlingh Onnes Laboratory of Leiden University,
see \cite{key-7,key-18,key-19} and references within. Spherical detectors
are important for the potential detection of the scalar component
of GWs that is admitted in ETG \cite{key-19}. In the case of interferometric
detectors, free falling masses are interferometer mirrors which can
be separated by kilometres (3km for Virgo, 4km for LIGO) \cite{key-1,key-7,key-18}.
In this way, GW tidal force is, in principle, several order of magnitude
larger than in bar detectors. Interferometers have very large bandwidth
($10-10000\mbox{ }Hz$) because mirrors are suspended to pendulums
having resonance in the Hz region. Thus, above such a resonance frequency,
mirrors work, in a good approximation, like freely falling masses
in the horizontal plane \cite{key-1,key-7,key-18}. 

Now, let us recall the importance to distinguish the gravitational
theories by using the observation of GWs \cite{key-20}. Motivations
to extend GTR arise from the fact that even though Einstein's theory
\cite{key-4} has achieved great success (see for example the opinion
of Landau, who said that GTR is, together with Quantum Field Theory,
the best scientific theory of all \cite{key-21}) and passed a lot
of experimental tests \cite{key-22} it has also showed some shortcomings
and flaws which today prompt theorists to ask if it is the definitive
theory of gravity \cite{key-3}. Differently from other field theories
like the electromagnetic theory, GTR is very difficult to quantize
\cite{key-23}. This fact rules out the possibility of treating gravitation
like other quantum theories, and precludes the unification of gravity
with other interactions. At the present time, it is not possible to
realize a consistent Quantum Gravity Theory which leads to the unification
of gravitation with the other forces \cite{key-23}. On the other
hand, one can define ETG, those semi-classical theories where the
Lagrangian is modified, with respect to the standard Einstein\textendash{}Hilbert
gravitational Lagrangian, adding high order terms to the curvature
invariants (terms like $R^{2}$, $R^{ab}R_{ab}$, $R^{abcd}R_{abcd}$,
$R\Box R$, $R\Box^{k}R$, in the sense of the so-called $f(R)$ theories,
see the recent review \cite{key-24}) and/or terms with scalar fields
non-minimally coupled to geometry (terms like $\phi^{2}R$ in the
sense of the so-called Scalar-Tensor Theories \cite{key-25}, i.e.
generalizations of the Jordan-Fierz-Brans-Dicke theory of gravitation
\cite{key-26,key-27,key-28}). In general, one has to emphasize that
terms like those are present in all the approaches to performing the
unification between gravity and other interactions. In addition, from
a cosmological point of view, such modifications of GTR produce inflationary
frameworks, which are very important as they solve a lot of problems
of the Standard Universe Model \cite{key-12}. Note that we are not
saying that GTR is wrong. It is well known that, even in the context
of extended theories, GTR remains the most important part of the structure
\cite{key-3,key-24}. We are only trying to understand if weak modifications
on such a structure could be needed to solve some theoretical and
observing problems. 

In the general context of cosmological evidences, there are other
considerations which suggest an extension of GTR. As a matter of fact,
the accelerated expansion of the Universe, which is today observed,
shows that the cosmological dynamic is dominated by the so-called
dark energy, which gives a large negative pressure. This is the standard
picture, in which such a new ingredient is considered as a source
of the right side of the field equations. It should be some form of
unclustered non-zero vacuum energy which, together with the clustered
dark matter, drives the global dynamics. This is the so-called \textquotedblleft{}\emph{concordance
model}\textquotedblright{} ($\Lambda CDM$), which gives, in agreement
with the CMBR, LSS and SNeIa data, a good tapestry of today's observed
Universe, but presents several shortcomings, such as the well-known
\textquotedblleft{}\emph{coincidence}\textquotedblright{} and \textquotedblleft{}\emph{cosmological
constant}\textquotedblright{} problems \cite{key-29}. An alternative
approach is to change the left side of the field equations, seeing
if observed cosmic dynamics can be achieved by extending GTR \cite{key-3}.
In this different context, we are not required to find candidates
for dark energy and dark matter, which till now have not been found,
but only the \textquotedblleft{}\emph{observed}\textquotedblright{}
ingredients, which are curvature and baryon matter, have to be taken
into account. Considering this point of view, one can think that gravity
is different at various scales and room for alternative theories is
present \cite{key-3,key-24}. In principle, the most popular dark
energy and dark matter models can be achieved considering $f(R)$
theories of gravity \cite{key-24}, where $R$ is the Ricci curvature
scalar, and/or STG \cite{key-25}. 

Also the Tensor-Vector-Scalar Theory (TVST) has attracted considerable
attention as an alternative to GTR \cite{key-33}. TVST is proposed
as a relativistic theory of Modified Newtonian Dynamics (MOND) \cite{key-33},
and it reproduces MOND in the weak acceleration limit. 

Let us recall the previous studies of how to distinguish alternative
gravitational theories from GTR \cite{key-20}. For example, STG could
be distinguished from GTR with surface atomic line redshift \cite{key-30},
with GWs \cite{key-31,key-32}, while the TVST theory could be distinguished
from GTR with surface atomic line redshift \cite{key-33}, with Shapiro
delays of gravitational waves and photon or neutrino \cite{key-34},
with GWs \cite{key-35,key-36}, with the rotational effect \cite{key-37}.
The recent result \cite{key-3} has shown that, if advanced projects
on the detection of GWs improve their sensitivity, allowing the Scientific
Community to perform a GW astronomy, accurate angle- and frequency-dependent
response functions of interferometers for GWs arising from various
theories of gravity will permit to discriminate among GTR and ETG
in an definitive way. This ultimate test will work because standard
GTR admits only two polarizations for GWs, while in all ETG the polarizations
are, at least, three, see \cite{key-3} for details. 

Recently, starting from the analysis in Ref. \cite{key-38}, some
papers in the literature have shown the importance of the \emph{gravitomagnetic}
effects in the framework of the GWs detection \cite{key-7}, \cite{key-39}
- \cite{key-42}. In fact, the so-called \emph{magnetic} components
of GWs have to be taken into account in the context of the total response
functions of interferometers for GWs propagating from arbitrary directions,
\cite{key-7}, \cite{key-38} - \cite{key-42}. In a recent paper,
the \emph{magnetic} component has been extended to GWs arising from
STG too \cite{key-43}. In particular, in Ref. \cite{key-43} it has
been shown that if one neglects the \emph{magnetic} contribution considering
only the low-frequency approximation of the \emph{electric} contribution,
an important portion of the signal could be, in principle, lost in
the case of STG too, in total analogy with the standard case of GTR
\cite{key-7}, \cite{key-38} - \cite{key-42}. Then, it is clear
that the computation of a more precise response function for the \emph{magnetic}
contribution is important also in the framework of the possibility
to distinguish other gravitational theories from GTR. 

On the other hand, in \cite{key-43} an error was present in the fundamental
equations (20) of such a paper \cite{key-63}. That error was dragged
along all the computations in \cite{key-43} by enabling incorrect
geometric factors in the angular dependence of the response function.
In this paper the original error and the geometric factors in the
angular dependence are corrected in order to obtain the correct response
function for the \emph{magnetic} component of GWs in STG. 

Before starting the analysis, let us explain the meaning of what is
\emph{magnetic} and what is \emph{electric} among the components of
GWs \cite{key-22}. Following \cite{key-38}, let us consider the
analogy between the motion of free masses in the field of a GW and
the motion of free charges in the field of an electromagnetic wave.
A GW drives the masses in the plane of the wave-front and also, to
a smaller extent, back and forth in the direction of the propagation
of the wave. To describe this motion, the notion of \emph{electric}
and \emph{magnetic} components of the gravitational force due to a
GW can be introduced, as it has been discussed in \cite{key-7}, \cite{key-38}
- \cite{key-43}. The analogy is not perfect, but it shows some important
features of the phenomenon \cite{key-38}. In Refs. \cite{key-7},
\cite{key-38} - \cite{key-43} the positions and motion of free test
masses has been analysed in the local inertial reference frame associated
with one of the masses, i.e. the beam-splitter in the case of an interferometer.
It is well known that this choice of coordinate system is the closest
thing to the global Lorentzian coordinates that are normally used
in electrodynamics \cite{key-22}. The distinction among the \emph{electric}
and \emph{magnetic} components of motion, as well as it is compared
with electrodynamics, is particularly clear in this description \cite{key-7},
\cite{key-38} - \cite{key-43}. When one interacts with the detection
of GWs, the usually used equations, with the curvature tensor in them,
are only the zero-order approximation in terms of $L/\lambda$, where
$L$ is the length of the arms of the interferometer and $\lambda$
the wave-length of the propagating GW \cite{key-7}, \cite{key-38}
- \cite{key-43}. This approximation is sufficient for the description
of the \emph{electric} part of the motion, which concerns frequencies
of order hundreds Hz, but it results insufficient for the description
of the \emph{magnetic} part, which can concern frequencies of order
KHzs. In the next approximation, which is a first order in terms of
$L/\lambda$, the geodesic deviation equation includes the derivatives
of the curvature tensor, and this approximation is fully sufficient
for the description of the \emph{magnetic} force and \emph{magnetic}
component of motion. One understands that the component of motion
which is called, with some reservations, \emph{magnetic} represents
the finite-wavelength correction to the usual infinite-wavelength
approximation \cite{key-7}, \cite{key-38} - \cite{key-43}. 

From the analyses in \cite{key-7}, \cite{key-38} - \cite{key-42},
it resulted that such a \textit{magnetic} component becomes particularly
important in the high frequency portion of the frequency range of
ground based interferometers and in future space based interferometers
for GWs which arises from standard GTR. The analysis has been extended
to GWs arising from STG too in \cite{key-43}. After a review of some
important issues in Section 2, in this paper we re-analyse the \textit{magnetic}
component in the framework of STG from a different point of view and
we correct an original error in \cite{key-43}, which generated incorrect
geometric factors in the angular dependence, in order to obtain the
correct response function for the \emph{magnetic} component of GWs
in STG. After this, we also compute a more precise response function
which will show that if one neglects the \textit{magnetic} contribution
considering only the low-frequency approximation of the \textit{electric}
contribution, an important portion of the signal, which could arrive
to about the $15\%$ for particular directions of the propagating
GWs, could be, in principle, lost.

It is important to discuss the splitting between \textit{magnetic}
and \emph{electric} components from another point of view \cite{key-44}.
In GTR, GWs are pure spin-2 tensor waves. In alternative theories
there can be other spin contributions to the field, and the waves
\cite{key-44}. In the particular case of this paper, which regards
STG, there is an additional scalar sector to the gravitational field,
responsible for a scalar sector to gravitational radiation. More specifically,
one may mathematically break the gravitational field in GTR between
\emph{electric-like} and \emph{magnetic-like} sectors, so called because
of formal mathematical similarities to their name sakes in Maxwell's
theory \cite{key-44}. This division of the full gravitational field
is most elegantly done in GTR using the Weyl tensor \cite{key-44,key-45,key-46}.
For a sake of completeness, this important point will be reviewed
in next Subsection 2.1.

At the end of the paper an expansion of the main results is also shown
in order to recall the presence of the magnetic component in GRT too
\cite{key-44}.

\section{A review of some important issues}

\subsection{Decomposition of the Weyl tensor into the electric and magnetic components}

In this Subsection, where we closely follow \cite{key-46}, we show
an irreducible splitting into \emph{electric} and \emph{magnetic}
parts for the Weyl tensor. 

Tidal forces in metric theories of gravity like GRT and STG are described
in a covariant way by the \emph{geodesic deviation equation} \cite{key-45,key-46,key-47}

\begin{equation}
\frac{D^{2}\xi^{a}}{ds^{2}}=-R_{mbn}^{\quad a}\frac{dx^{m}}{ds}\frac{dx^{n}}{ds}\xi^{b},\label{eq: def geo}\end{equation}

where $\xi^{a}$ is the separation vector between two test masses
\cite{key-45,key-46,key-47}, i.e. \begin{equation}
\xi^{b}\equiv x_{m1}^{b}-x_{m2}^{b},\label{eq: diff}\end{equation}
 $\frac{D}{ds}$ is the covariant derivative and $s$ the affine parameter
along a geodesic \cite{key-45,key-46,key-47}. In this paper Latin
indices are used for 4-dimensional quantities, Greek indices for 3-dimensional
ones and the author works with $G=1$, $c=1$ and $\hbar=1$ (natural
units). Eq. (\ref{eq: diff}) gives the relative acceleration of two
neighbouring particles with the same 4-velocity $\frac{dx^{a}}{ds}$.
If one wants to find the electromagnetic analogue to (\ref{eq: def geo}),
a very intrinsic difference between the two interactions has to be
recalled. While the ratio between gravitational and inertial mass
is universal, the same does not apply to the ratio between electrical
charge and inertial mass. In other words, there is no electromagnetic
counterpart of the equivalence principle \cite{key-46}. Thus, the
analogue electromagnetic problem will consist in considering two neighbouring
particles with the same 4-velocity $\frac{dx^{a}}{ds}$ in an electromagnetic
field on a flat Minkowskian spacetime, by assuming the extra condition
that the two particles have the same $q/m$ ratio \cite{key-46}.
Under these constrains, one obtains the worldline deviation equation
as \cite{key-46} 

\begin{equation}
\frac{D^{2}\xi^{a}}{ds^{2}}=\frac{q}{m}F_{m;b}^{a}\frac{dx^{m}}{ds}\xi^{b},\label{eq: wordline}\end{equation}

where $F_{ba}^{d}$ is the electromagnetic tensor \cite{key-21}.
By comparing (\ref{eq: def geo}) with (\ref{eq: wordline}) one gets
a physical analogy between the two tensors \cite{key-46}: 

\begin{equation}
E_{ab}^{gravity}\equiv R_{ambn}\frac{dx^{m}}{ds}\frac{dx^{n}}{ds}\mbox{ }\leftrightarrow E_{ab}\equiv F_{am;b}\frac{dx^{m}}{ds}.\label{eq: tensori elettrici}\end{equation}

The tensor $E_{ab}$ is the covariant derivative of the electric field,
which is defined like $E^{a}\equiv F^{ab}\frac{dx_{b}}{ds}$ and it
is seen by an observer having a 4-velocity $\frac{dx^{a}}{ds}.$ It
is usually called the \emph{electric tidal tensor}. The gravitational
counterpart $E_{ab}^{gravity}$ is usually called the \emph{electric
gravitational tidal tensor}. The different signs in (\ref{eq: def geo})
and (\ref{eq: wordline}) are due by the different interaction (attractive
or repulsive) between masses or charges of the same sign \cite{key-46}.
In analogous way one defines the \emph{magnetic tidal tensor} as 

\begin{equation}
B_{ab}\equiv\star F_{am;b}\frac{dx^{m}}{ds}=\frac{1}{2}\epsilon_{am}^{cl}F_{cl,b}\frac{dx^{m}}{ds},\label{eq: magnetic tidal tensor}\end{equation}

where $\epsilon_{abcd}$ is the Levi-Civita tensor and $\star$ denotes
the Hodge dual \cite{key-46}. $B_{ab}$ represents the tidal effects
produced by the magnetic field which is defined like $B^{a}\equiv\star F^{ab}\frac{dx_{b}}{ds}$,
seen by an observer who has a 4-velocity $\frac{dx^{c}}{ds}.$

Then, by working with the Riemann tensor, one introduces the so called
\emph{magnetic part of the Riemann tensor}:

\begin{equation}
B_{ab}^{grav}\equiv\star R_{ambn}\frac{dx^{m}}{ds}\frac{dx^{n}}{ds}=\frac{1}{2}\epsilon_{am}^{cl}R_{clbn}\frac{dx^{m}}{ds}\frac{dx^{n}}{ds},\label{eq: magnetic part of the Riemann tensor}\end{equation}

which is the the physical gravitational analogue of $B_{ab}$ and
is usually called the \emph{magnetic gravitational tidal tensor} \cite{key-46}.

Now, let us introduce the decomposition of the Riemann tensor \cite{key-46,key-48}

\begin{equation}
R_{abcd}=C_{abcd}+g_{a[c}R_{d]b}+g_{b[d}R_{c]a}+\frac{1}{3}g_{a[d}g_{c]b}R,\label{eq: decomposition}\end{equation}

where $C_{abcd}$ is the Weyl tensor. Like the Riemann curvature tensor,
the Weyl tensor expresses the tidal force that a body feels when moving
along a geodesic \cite{key-48}. The Weyl tensor differs from the
Riemann curvature tensor in that it does not convey information on
how the volume of the body changes, but rather only how the shape
of the body is distorted by the tidal force \cite{key-48}. The Weyl
tensor is traceless and shows the property \cite{key-46,key-48}

\begin{equation}
\star C_{abcd}=C\star_{abcd}.\label{eq: duale}\end{equation}

By introducing the electric and magnetic parts of the Weyl tensor,
both of which are symmetric and traceless, i.e. \cite{key-46}

\begin{equation}
\varepsilon_{ab}\equiv C_{acbn}\frac{dx^{c}}{ds}\frac{dx^{n}}{ds},\mbox{ }H_{ab}\equiv\star C_{acbn}\frac{dx^{c}}{ds}\frac{dx^{n}}{ds},\label{eq: Weyl}\end{equation}

$E_{ab}^{gravity}$ and $B_{ab}^{gravity}$ read \cite{key-46}

\begin{equation}
\begin{array}{c}
E_{ab}^{gravity}=\varepsilon_{ab}+\frac{1}{2}[g_{ab}R_{cd}\frac{dx^{c}}{ds}\frac{dx^{d}}{ds}+\\
\\-R_{ab}-2\frac{dx_{(a}}{ds}R_{b)d}\frac{dx^{d}}{ds}]+\frac{1}{6}R[g_{ab}+\frac{dx_{a}}{ds}\frac{dx_{b}}{ds}]\end{array}\label{eq: tensore elettrico ri-definito}\end{equation}

and 

\begin{equation}
B_{ab}^{gravity}=\mbox{ }H_{ab}+\frac{1}{2}\epsilon_{abnc}R_{d}^{n}\frac{dx^{c}}{ds}\frac{dx^{d}}{ds}.\label{eq: tensore magnetico ri-definito}\end{equation}

These expressions can be used to obtain the gravitational analogue
of Maxwell equations, see \cite{key-46} for details.

\subsection{The linearized Scalar-Tensor Gravity}

The most general action of STG in four dimensions is given by \cite{key-19,key-25,key-43,key-47,key-52}

\begin{equation}
S=\int d^{4}x\sqrt{-g}[f(\phi)R+\frac{1}{2}g^{mn}\phi_{;m}\phi_{;n}-V(\phi)+\mathcal{L}_{Mass-Energy}].\label{eq: scalar-tensor}\end{equation}

Choosing

\begin{equation}
\begin{array}{ccc}
\varphi=f(\phi) & \omega(\varphi)=\frac{f(\phi)}{2f'(\phi)} & W(\varphi)=V(\phi(\varphi))\end{array}\label{eq: scelta}\end{equation}

Eq. (\ref{eq: scalar-tensor}) reads

\begin{equation}
S=\int d^{4}x\sqrt{-g}[\varphi R-\frac{\omega(\varphi)}{\varphi}g^{mn}\varphi_{;m}\varphi_{;n}-W(\varphi)+\mathcal{L}_{Mass-Energy}],\label{eq: scalar-tensor2}\end{equation}

which is a generalization of the Jordan-Fierz-Brans-Dicke theory of
gravitation \cite{key-26,key-27,key-28}.

By varying the action (\ref{eq: scalar-tensor2}) with respect to
$g_{mn}$ and to the scalar field $\varphi$ the field equations are
obtained \cite{key-19,key-25,key-43,key-47,key-52}: \begin{equation}
\begin{array}{c}
G_{mn}=-\frac{4\pi\tilde{G}}{\varphi}T_{mn}^{(Mass-Energy)}+\frac{\omega(\varphi)}{\varphi^{2}}(\varphi_{;m}\varphi_{;n}-\frac{1}{2}g_{mn}g^{ab}\varphi_{;a}\varphi_{;b})+\\
\\+\frac{1}{\varphi}(\varphi_{;mn}-g_{mn}\square\varphi)+\frac{1}{2\varphi}g_{mn}W(\varphi),\end{array}\label{eq: einstein-general}\end{equation}
with associated a Klein - Gordon equation for the scalar field

\begin{equation}
\square\varphi=\frac{1}{2\omega(\varphi)+3}(-4\pi\tilde{G}T^{(Mass-Energy)}+2W(\varphi)+\varphi W'(\varphi)+\frac{d\omega(\varphi)}{d\varphi}g^{mn}\varphi_{;m}\varphi_{;n}.\label{eq: KG}\end{equation}

In the above equations $T_{mn}^{(Mass-Energy)}$ is the ordinary stress-energy
tensor of the matter and $\tilde{G}$ is a dimensional, strictly positive,
constant. The Newton constant is replaced by the effective coupling

\begin{equation}
G_{eff}=-\frac{1}{2\varphi},\label{eq: newton eff}\end{equation}

which is, in general, different from $G$. GTR is re-obtained when
the scalar field coupling is 

\begin{equation}
\varphi=const.=-\frac{1}{2}.\label{eq: varphi}\end{equation}

To study GWs, the linearized theory in vacuum ($T_{mn}^{(Mass-Energy)}=0$)
with a little perturbation of the background has to be analysed. The
background is assumed given by the Minkowskian background plus $\varphi=\varphi_{0}$
and $\varphi_{0}$ is also assumed to be a minimum for $W$ \cite{key-19,key-47}: 

\begin{equation}
W\simeq\frac{1}{2}\alpha\delta\varphi^{2}\Rightarrow W'\simeq\alpha\delta\varphi\label{eq: minimo}\end{equation}

Putting

\begin{equation}
\begin{array}{c}
g_{mn}=\eta_{mn}+h_{mn}\\
\\\varphi=\varphi_{0}+\delta\varphi.\end{array}\label{eq: linearizza}\end{equation}

and, to first order in $h_{mn}$ and $\delta\varphi$, if one calls
$\widetilde{R}_{mnrs}$ , $\widetilde{R}_{mn}$ and $\widetilde{R}$
the linearized quantity which correspond to $R_{mnrs}$ , $R_{\mu\nu}$
and $R$, the linearized field equations are obtained \cite{key-19,key-47}:

\begin{equation}
\begin{array}{c}
\widetilde{R}_{mn}-\frac{\widetilde{R}}{2}\eta_{mn}=-\partial_{m}\partial_{n}\Phi+\eta_{mn}\square\Phi\\
\\{}\square\Phi=m^{2}\Phi,\end{array}\label{eq: linearizzate1}\end{equation}

where

\begin{equation}
\begin{array}{c}
\Phi\equiv-\frac{\delta\varphi}{\varphi_{0}}\\
\\m^{2}\equiv\frac{\alpha\varphi_{0}}{2\omega+3}.\end{array}\label{eq: definizione}\end{equation}

The case in which it is $\omega=const.$ and $W=0$ in Eqs. (\ref{eq: einstein-general})
and (\ref{eq: KG}) has been analysed in \cite{key-19,key-47} with
a treatment which generalized the {}``canonical'' linearization
of GTR \cite{key-22}.

For a sake of completeness, let us complete the linearization process
by closely following \cite{key-19,key-47}.

The linearized field equations become

\begin{equation}
\begin{array}{c}
\widetilde{R}_{mn}-\frac{\widetilde{R}}{2}\eta_{mn}=\partial_{m}\partial_{n}\xi+\eta_{mn}\square\Phi\\
\\\square\Phi=0\end{array}\label{eq: linearizzate2}\end{equation}

Let us put 

\begin{equation}
\begin{array}{c}
\bar{h}_{mn}\equiv h_{mn}-\frac{h}{2}\eta_{mn}+\eta_{mn}\Phi\\
\\\bar{h}\equiv\eta^{mn}\bar{h}_{mn}=-h+4\Phi,\end{array}\label{eq: h barra}\end{equation}

with $h\equiv\eta^{mn}h_{mn}$, where the inverse transform is the
same

\begin{equation}
\begin{array}{c}
h_{mn}=\bar{h}_{mn}-\frac{\bar{h}}{2}\eta_{mn}+\eta_{mn}\Phi\\
\\h=\eta^{mn}h_{mn}=-\bar{h}+4\Phi.\end{array}\label{eq: h}\end{equation}

By putting the first of Eqs. (\ref{eq: h}) in the first of the field
Eqs. (\ref{eq: linearizzate2}) it is

\begin{equation}
\square\bar{h}_{mn}-\partial_{m}(\partial^{a}\bar{h}_{an})-\partial_{n}(\partial^{a}\bar{h}_{an})+\eta_{mn}\partial^{b}(\partial^{a}\bar{h}_{ab}).\label{eq: onda}\end{equation}

Now, let us consider the gauge transform (Lorenz condition)

\begin{equation}
\begin{array}{c}
\bar{h}_{mn}\rightarrow\bar{h}'_{mn}=\bar{h}_{mn}-\partial_{(m}\epsilon_{n)}+\eta_{mn}\partial^{a}\epsilon_{a}\\
\\\bar{h}\rightarrow\bar{h}'=\bar{h}+2\partial^{a}\epsilon_{a}\\
\\\Phi\rightarrow\Phi'=\Phi\end{array}\label{eq: gauge lorenzt}\end{equation}

with the condition $\square\epsilon_{n}=\partial^{m}\bar{h}_{mn}$
for the parameter $\epsilon^{m}$. It is 

\begin{equation}
\partial^{m}\bar{h}'_{mn}=0,\label{eq: cond lorentz}\end{equation}

and, omitting the $'$, the field equations can be rewritten like

\begin{equation}
\square\bar{h}_{mn}=0\label{eq: onda T}\end{equation}

\begin{equation}
\square\Phi=0;\label{eq: onda S}\end{equation}

solutions of Eqs. (\ref{eq: onda T}) are plan waves:

\begin{equation}
\bar{h}_{mn}=A_{mn}(\overrightarrow{k})\exp(ik^{a}x_{a})+c.c.\label{eq: sol T}\end{equation}

\begin{equation}
\Phi=a(\overrightarrow{k})\exp(ik^{a}x_{a})+c.c.\label{eq: sol S}\end{equation}

Thus, Eqs. (\ref{eq: onda T}) and (\ref{eq: sol T}) are the equation
and the solution for the tensor waves exactly like in GTR \cite{key-22},
while Eqs. (\ref{eq: onda S}) and (\ref{eq: sol S}) are respectively
the equation and the solution for the scalar mode.

The solutions (\ref{eq: sol T}) and (\ref{eq: sol S}) preserve the
conditions

\begin{equation}
\begin{array}{c}
k^{a}k_{a}=0\\
\\k^{m}A_{mn}=0,\end{array}\label{eq: vincoli}\end{equation}

which arises respectively from the field equations and from Eq. (\ref{eq: cond lorentz}).

The first of Eqs. (\ref{eq: vincoli}) shows that perturbations have
the speed of light, the second the transversal effect of the field.

Fixed the Lorenz gauge, another transformation with $\square\epsilon^{m}=0$
can be made; let us take

\begin{equation}
\begin{array}{c}
\square\epsilon^{m}=0\\
\\\partial_{m}\epsilon^{m}=-\frac{\bar{h}}{2}+\Phi,\end{array}\label{eq: gauge2}\end{equation}

which is permitted because $\square\Phi=0=\square\bar{h}$. We obtain

\begin{equation}
\begin{array}{ccc}
\bar{h}=2\Phi & \Rightarrow & \bar{h}_{mn}=h_{mn}\end{array},\label{eq: h ug h}\end{equation}

i.e. $h_{mn}$ is a transverse plane wave too. The gauge transformations

\begin{equation}
\begin{array}{c}
\square\epsilon^{m}=0\\
\\\partial_{\mu}\epsilon^{m}=0,\end{array}\label{eq: gauge3}\end{equation}
preserve the conditions

\begin{equation}
\begin{array}{c}
\partial^{m}\bar{h}_{mn}=0\\
\\\bar{h}=2\Phi.\end{array}\label{eq: vincoli 2}\end{equation}

Considering a wave propagating in the positive $z$ direction 

\begin{equation}
k^{m}=(k,0,0k),\label{eq: k}\end{equation}

the second of Eqs. (\ref{eq: vincoli}) implies

\begin{equation}
\begin{array}{c}
A_{0n}=-A_{3n}\\
\\A_{n0}=-A_{n3}\\
\\A_{00}=-A_{30}+A_{33}.\end{array}\label{eq: A}\end{equation}

Now, let us see the freedom degrees of $A_{mn}$. We was started with
10 components ($A_{mn}$ is a symmetric tensor); 3 components have
been lost for the transversal condition, more, the condition (\ref{eq: h ug h})
reduces the component to 6. One can take $A_{00}$, $A_{11}$, $A_{22}$,
$A_{21}$, $A_{31}$, $A_{32}$ like independent components; another
gauge freedom can be used to put to zero three more components (i.e.
only three of $\epsilon^{\mu}$ can be chosen, the fourth component
depends from the others by $\partial_{m}\epsilon^{m}=0$).

Then, by taking 

\begin{equation}
\begin{array}{c}
\epsilon_{m}=\tilde{\epsilon}_{m}(\overrightarrow{k})\exp(ik^{a}x_{a})+c.c.\\
\\k^{m}\tilde{\epsilon}_{m}=0,\end{array}\label{eq: ancora gauge}\end{equation}

the transform law for $A_{mn}$ is (see Eqs. (\ref{eq: gauge lorenzt})
and (\ref{eq: sol T}))

\begin{equation}
A_{mn}\rightarrow A'_{mn}=A_{mn}-2ik(_{m}\tilde{\epsilon}_{n}).\label{eq: trasf. tens.}\end{equation}

Thus, for the six components of interest

\begin{equation}
\begin{array}{ccc}
A_{00} & \rightarrow & A_{00}+2ik\tilde{\epsilon}_{0}\\
A_{11} & \rightarrow & A_{11}\\
A_{22} & \rightarrow & A_{22}\\
A_{21} & \rightarrow & A_{21}\\
A_{31} & \rightarrow & A_{31}-ik\tilde{\epsilon}_{1}\\
A_{32} & \rightarrow & A_{32}-ik\tilde{\epsilon}_{2}.\end{array}\label{eq: sei tensori}\end{equation}

The physical components of $A_{mn}$ are the gauge-invariants $A_{11}$,
$A_{22}$ and $A_{21}$, thus one can chose $\tilde{\epsilon}_{n}$
to put equal to zero the others.

The scalar field is obtained by Eq. (\ref{eq: h ug h}):

\begin{equation}
\bar{h}=h=h_{11}+h_{22}=2\Phi.\label{eq: trovato scalare}\end{equation}

In this way, the total perturbation of a GW propagating in the $z-$
direction in this gauge is

\begin{equation}
h_{mn}(t+z)=A^{+}(t+z)e_{mn}^{(+)}+A^{\times}(t+z)e_{mn}^{(\times)}+\Phi(t+z)e_{mn}^{(s)}.\label{eq: perturbazione totale}\end{equation}

The term $A^{+}(t+z)e_{mn}^{(+)}+A^{\times}(t+z)e_{mn}^{(\times)}$
describes the two standard (i.e. tensor) polarizations of GWs which
arises from GTR in the TT gauge \cite{key-22}, while the term $\Phi(t+z)e_{mn}^{(s)}$
is a third polarization which is due by the extension of the TT gauge
to the STG case.

For a purely scalar GW the metric perturbation (\ref{eq: perturbazione totale})
reduces to 

\begin{equation}
h_{mn}=\Phi e_{mn}^{(s)},\label{eq: perturbazione scalare}\end{equation}

and the correspondent line element is \cite{key-19,key-47} \begin{equation}
ds^{2}=dt^{2}-dz^{2}-(1+\Phi)dx^{2}-(1+\Phi)dy^{2},\label{eq: metrica TT scalare}\end{equation}
with $\Phi=\Phi_{0}e^{i\omega(t+z)}.$

The wordlines $x,y,z=const.$ are timelike geodesics representing
the histories of free test masses, see the analogy with tensor waves
in \cite{key-22}.

\subsection{Quadrupole, dipole and monopole modes: potential detection}

It is important to recall that in the case of STG the scalar GWs will
be excited as well as tensor GWs, thus, in principle, the promising
GW sources of scalar GWs and their frequencies are exactly the same
of ordinary tensor GW. In fact, the production of scalar gravitational
radiation is no different than the production of any other type of
radiation \cite{key-65}. If one wants to produce electromagnetic
radiation at, say, 1 KHz, one needs to take electric charges and vibrate
them at 1 KHz \cite{key-65}. The same holds for both of tensor and
scalar gravitational radiation; waves of a certain frequency are produced
when the characteristic time for the matter and energy in the universe
to shift about is comparable to the period of the waves \cite{key-65}.
Coalescing binaries systems emit at frequencies around 1 KHz \cite{key-1},
while single rotating pulsars have a spin frequency which lies in
the hectohertz \emph{{}``sweet spot}'' of current detectors, i.e.
at order hundreds Hz \cite{key-49}. The frequency of GW emission
from collapsed objects like Supernovae is in the range 50Hz to a few
KHz \cite{key-50}. The stochastic background of GWs has spectrum
which is flat along the frequency range $10^{-16}\leq f\leq10^{8}\mbox{ }Hz$
\cite{key-51}. 

An important difference with respect to standard GTR is that the scalar
GWs will radiate even in the case that the event would be spherically
symmetric \cite{key-20}. Thus, we understand that in the case of
almost spherically symmetric events the energy emitted by tensor modes
can be neglected \cite{key-47,key-54} (in the sense that the scalar
modes largely exceed the tensor ones). Let us see this issue in detail.

We emphasize that in this Subsection we closely follow the papers
\cite{key-47,key-53,key-54}. 

In the framework of GWs, the more important difference between GTR
and STG is the existence, in the latter, of dipole and monopole radiation
\cite{key-47,key-53}. In GTR, for slowly moving systems, the more
important multi-pole contribution to gravitational radiation is the
quadrupole one. The result is that the dominant radiation-reaction
effects are at order $(\frac{v}{c})^{5}$, where $v$ is the orbital
velocity. The rate, due to quadrupole radiation, at which a binary
system loses energy is given, in GTR, by \cite{key-47,key-53}

\begin{equation}
(\frac{dE}{dt})_{quadrupole}=-\frac{8}{15}\eta^{2}\frac{m^{4}}{r^{4}}(12v^{2}-11\dot{r}^{2}).\label{eq:  Will}\end{equation}

$\eta$ and $m$ are, respectively, the reduced mass parameter and
total mass, given by $\eta=\frac{m_{1}m_{2}}{(m_{1}+m_{2})^{2}}$
, and $m=m_{1}+m_{2}$ .

$r,$ $v,$ and $\dot{r}$ represent, respectively, the orbital separation,
relative orbital velocity, and radial velocity.

In STG, Eq. (\ref{eq:  Will}) is modified by corrections to the coefficients
of $O(\frac{1}{\omega})$, where $\omega$ is the Brans-Dicke parameter
(STG also predicts monopole radiation, but in binary systems it contributes
only to these $O(\frac{1}{\omega})$ corrections) \cite{key-47,key-53}.
The important modification in STG is the additional energy loss caused
by dipole modes. By analogy with electrodynamics, dipole radiation
is a $(v/c)^{3}$ effect, potentially much stronger than quadrupole
radiation. However, in STG, the gravitational \textquotedblleft{}\emph{dipole
moment}\textquotedblright{} is governed by the difference $s_{1}-s_{2}$
between the bodies, where $s_{i}$ is a measure of the self-gravitational
binding energy per unit rest mass of each body \cite{key-47,key-53}.
$s_{i}$ represents the \textquotedblleft{}\emph{sensitivity}\textquotedblright{}
of the total mass of the body to variations in the background value
of the Newton constant, which, in this theory, is a function of the
scalar field \cite{key-47,key-53}: 

\begin{equation}
s_{i}=\left(\frac{\partial(\ln m_{i})}{\partial(\ln G)}\right)_{N}.\label{eq: Will 2}\end{equation}

\emph{$G$} is the effective Newtonian constant at the star and the
subscript $N$ denotes holding baryon number fixed. 

Defining $S\equiv s_{1}-s_{_{2}}$, to first order in $\frac{1}{\omega}$
the energy loss caused by dipole radiation is given by \cite{key-47,key-53}
\begin{equation}
(\frac{dE}{dt})_{dipole}=-\frac{2}{3}\eta^{2}\frac{m^{4}}{r^{4}}(\frac{S^{2}}{\omega}).\label{eq:  Will 3}\end{equation}

In STG, the sensitivity of a black hole is always $s_{BH}=0.5$ \cite{key-47,key-53},
while the sensitivity of a neutron star varies with the equation of
state and mass. For example, $s_{NS}\approx0.12$ for a neutron star
of mass order $1.4M_{\circledcirc}$, being $M_{\circledcirc}$ the
solar mass \cite{key-47,key-53}. 

Binary black-hole systems are not at all promising for studying dipole
modes because $s_{BH1}-s_{BH2}=0,$ a consequence of the no-hair theorems
for black holes \cite{key-47,key-53}. In fact, black holes radiate
away any scalar field, so that a binary black hole system in STG behaves
as if GTR. Similarly, binary neutron star systems are also not effective
testing grounds for dipole radiation \cite{key-47,key-53}. This is
because neutron star masses tend to cluster around the Chandrasekhar
limit of $1.4M_{\circledcirc}$, and the sensitivity of neutron stars
is not a strong function of mass for a given equation of state. Thus,
in systems like the binary pulsar, dipole radiation is naturally suppressed
by symmetry, and the bound achievable cannot compete with those from
the solar system \cite{key-47,key-53}. Hence the most promising systems
are mixed: BH-NS, BH-WD, or NS-WD. 

The emission of monopole radiation from STG is very important in the
collapse of quasi-spherical astrophysical objects because in this
case the energy emitted by quadrupole modes can be neglected \cite{key-22,key-47,key-54}.
In \cite{key-54} it has been shown that, in the formation of a neutron
star, monopole waves interact with the detectors as well as quadrupole
ones. In that case, the field-dependent coupling strength between
matter and the scalar field has been assumed to be a linear function.
In the notation of this paper such a coupling strength is given by
$\alpha\equiv\frac{1}{2\omega(\varphi)+3}$ in Eq. (\ref{eq: KG}).
Then \cite{key-54}

\begin{equation}
\alpha=\alpha_{0}+\beta_{0}(\varphi-\varphi_{0})\label{eq: accoppiamento}\end{equation}

and the amplitude of the scalar polarization results \cite{key-54}

\begin{equation}
\Phi\propto\frac{\alpha_{0}}{d}\label{eq: ampiezza da supernova}\end{equation}

where $d$ is the distance of the collapsing neutron star expressed
in meters. 

On the other hand, such signals will be quite weak. Let us discuss
the experimental sensitivity required to detect them. We have also
to compare with the sensitivities of ongoing and future experiments.
To make this, we consider an astrophysical event which produces GWs
and which can, in principle, help to simplify the problem. In previous
discussion we analysed two potential sources of potential detectable
scalar radiation:
\begin{enumerate}
\item mixed binary systems like BH-NS, BH-WD, or NS-WD;
\item the gravitational collapse of quasi-spherical astrophysical objects.
\end{enumerate}
The second source looks propitious because in such a case the energy
emitted by quadrupole modes can be neglected \cite{key-47,key-54}
(in the sense that the monopole modes largely exceed the quadrupole
ones. In fact, if the collapse is completely spherical, the quadrupole
modes are totally removed \cite{key-22}). In that case, only the
motion of the test masses due to the scalar component has to be analysed. 

The authors of \cite{key-54} analysed the interesting case of the
formation of a neutron star through a gravitational collapse. In that
case, they found that a collapse occurring closer than 10 kpc from
us (half of our Galaxy) needs a sensitivity of  $3*10^{-23}\mbox{ }\sqrt{Hz}$
at $800\mbox{ }Hz$ (which is the characteristic frequency of such
events) to potential detect the strain which is generated by the scalar
component in the arms of LIGO. 

At the present time, the sensitivity of LIGO at about $800\mbox{ }Hz$
is $10^{-22}\mbox{ }\sqrt{Hz}$ while the sensitivity of the Enhanced
LIGO Goal is predicted to be $8*10^{-22}\mbox{ }\sqrt{Hz}$ at $800\mbox{ }Hz$
\cite{key-1}. Then, for a potential detection of the scalar mode
we have to hope in Advanced LIGO Baseline High Frequency and/or in
Advanced LIGO Baseline Broadband. In fact, the sensitivity of these
two advanced configuration is predicted to be $6*10^{-23}\mbox{ }\sqrt{Hz}$
at $800\mbox{ }Hz$ \cite{key-1}.

Another clarification is needed on the potential detection of the
scalar mode \cite{key-20}. To identify the scalar GW, one needs to
prepare several detectors \cite{key-20}. In fact, detectors to be
cross-correlated must be, at least two \cite{key-19,key-64}. A cross-correlation
can concern two different interferometers, like discussed for example
in \cite{key-64} or, alternatively, an interferometer can be cross-correlated
with a resonance bar \cite{key-19}. In \cite{key-19} the interesting
case of the cross-correlation between the Virgo interferometer and
the monopole mode of the MiniGRAIL resonant sphere for the detection
of the scalar mode has been analysed. Even if such a cross correlation
is very small, it has been shown that a maximum is present at about
$2710\mbox{ }Hz$, i.e. within the sensitivity's range of both of
MiniGRAIL and Virgo \cite{key-19}. Then, if the eventual detection
of a monopole mode of the MiniGRAIL bar at about $2710\mbox{ }Hz$
will coincide with a signal detected by the Virgo interferometer at
the same frequency, such a detection will be a strong endorsement
for Scalar Tensor Theories of Gravity. Indeed, the monopole mode of
a sphere cannot be excited by ordinary tensor waves arising from standard
GR, see \cite{key-19} for details.

\subsection{A note on conformal frames}

Concerning scalar GWs it is important clarify that the results in
Einstein frame will \emph{not} be same as those in physical frame
(Jordan-Fierz-Brans-Dicke frame) \cite{key-20}.

The author recently discussed this important issue in Ref. \cite{key-47}.
The key point is that the motion in the Einstein frame \emph{is not
geodesic} \cite{key-47,key-55,key-56}, and this point strongly endorses
deviations from equivalence principle and non-metric gravity theories
in the Einstein frame \cite{key-47,key-55,key-56}. The author showed
in \cite{key-47} that the geodesic deviation equation (\ref{eq: def geo}),
which governs GWs signals in the gauge of the local observer, changes
in the conformal Einstein frame becoming \cite{key-47}

\begin{equation}
\frac{D^{2}\xi^{d}}{ds^{2}}=-\tilde{R}_{abc}^{\quad d}\frac{dx^{c}}{ds}\frac{dx^{b}}{ds}\xi^{a}-\sqrt{\frac{4\pi}{|2\omega+3|}}\frac{D}{ds}(\partial^{d}\tilde{\varphi}),\label{eq: geo 3}\end{equation}

where $\tilde{R}_{abc}^{\quad d}$ is the rescaled Riemann tensor
in the conformal Einstein frame \cite{key-47,key-56}. Thus, an extra
term of the geodesic deviation equations, which is not present in
the Jordan frame, see Eq. (\ref{eq: def geo}), is present in the
Einstein frame, i.e. the term $-\sqrt{\frac{4\pi}{|2\omega+3|}}\frac{D}{ds}(\partial^{d}\tilde{\varphi})$
\cite{key-47}. This key point implies that the motion of the test
masses due to the scalar component of GWs in STG is \emph{different}
in the two frames. Such a motion has been carefully examined, in both
of the two frames, at first order in the geodesic deviation in Ref.
\cite{key-47}.

\section{\emph{Electric} and \emph{magnetic} components}

In a laboratory environment on Earth, the coordinate system in which
the space-time is locally flat is typically used \cite{key-22} and
the distance between any two points is given simply by the difference
in their coordinates in the sense of Newtonian physics. In this frame,
called the frame of the local observer, scalar GWs manifest themselves
by exerting tidal forces on the masses (the mirror and the beam-splitter
in the case of an interferometer). 

A detailed analysis of the frame of the local observer is given in
Ref. \cite{key-22}, sect. 13.6. Here only the more important features
of this frame are resumed:

the time coordinate $x_{0}$ is the proper time of the observer O;

spatial axes are centred in O;

in the special case of zero acceleration and zero rotation the spatial
coordinates $x_{j}$ are the proper distances along the axes and the
frame of the local observer reduces to a local Lorentz frame: in this
case the line element reads 

\begin{equation}
ds^{2}=-(dx^{0})^{2}+\delta_{\mu\nu}dx^{\mu}dx^{\nu}+O(|x^{j}|^{2})dx^{a}dx^{b};\label{eq: metrica local lorentz}\end{equation}

the effect of GWs on test masses is described by the equation for
geodesic deviation in this frame

\begin{equation}
\ddot{x^{\mu}}=-\widetilde{R}_{0\nu0}^{\mu}x^{\nu},\label{eq: deviazione geodetiche}\end{equation}
where $\widetilde{R}_{0\nu0}^{\mu}$ are the components of the linearized
Riemann tensor \cite{key-22}. 

Labelling the coordinates of the TT gauge with $t_{tt},x_{tt},y_{tt},z_{tt}$,
in \cite{key-43}, the coordinate transformation $x^{a}=x^{a}(x_{tt}^{b})$
from the TT coordinates to the frame of the local observer was written
as (Eqs. 20 in \cite{key-43})

\begin{equation}
\begin{array}{c}
t=t_{tt}+\frac{1}{4}(x_{tt}^{2}-y_{tt}^{2})\dot{\Phi}\\
\\x=x_{tt}+\frac{1}{2}x_{tt}\Phi+\frac{1}{2}x_{tt}z_{tt}\dot{\Phi}\\
\\y=y_{tt}+\frac{1}{2}y_{tt}\Phi+\frac{1}{2}y_{tt}z_{tt}\dot{\Phi}\\
\\z=z_{tt}-\frac{1}{4}(x_{tt}^{2}-y_{tt}^{2})\dot{\Phi},\end{array}\label{eq: trasf. coord. errata}\end{equation}

where it is $\dot{\Phi}\equiv\frac{\partial\Phi}{\partial t}$, see
the analogy with tensor waves of standard General Relativity in \cite{key-38,key-39,key-40,key-41,key-42}.
But we have to emphasize that in Eq. (\ref{eq: trasf. coord. errata})
an error is present \cite{key-63}. In fact, the extra (scalar) polarization
in Eq. (46) is symmetric with respect to rotations around the z-axis
\cite{key-63}. Therefore, the z-displacement of a test particle can
depend on its radial coordinate in xy-plane, but not on the positional
angle in this plane \cite{key-63}. However, such a positional dependence
is implied by the combination of $x_{tt}$ and $y_{tt}$ factors in
the last line of Eq. (55) \cite{key-63}. This line cannot be correct
\cite{key-63}. Clearly, the error is the sign minus before $y_{tt}^{2}$
in both of the first and the last lines of Eq. (55). Thus, the correct
coordinate transformation from the TT coordinates to the frame of
the local observer is \begin{equation}
\begin{array}{c}
t=t_{tt}+\frac{1}{4}(x_{tt}^{2}+y_{tt}^{2})\dot{\Phi}\\
\\x=x_{tt}+\frac{1}{2}x_{tt}\Phi+\frac{1}{2}x_{tt}z_{tt}\dot{\Phi}\\
\\y=y_{tt}+\frac{1}{2}y_{tt}\Phi+\frac{1}{2}y_{tt}z_{tt}\dot{\Phi}\\
\\z=z_{tt}-\frac{1}{4}(x_{tt}^{2}+y_{tt}^{2})\dot{\Phi},\end{array}\label{eq: trasf. coord.}\end{equation}

which respects the symmetry with respect to rotations around the z-axis
of the third scalar polarization. The coefficients of this transformation
(components of the metric and its first time derivative) are taken
along the central wordline of the local observer \cite{key-43}. The
linear and quadratic terms, as powers of $x_{tt}^{a}$, are unambiguously
determined by the conditions of the frame of the local observer, while
the cubic and higher-order corrections are not determined by these
conditions \cite{key-38,key-39,key-40,key-41,key-42,key-43}. 

Considering a free mass riding on a timelike geodesic ($x=l_{1}$,
$y=l_{2},$ $z=l_{3}$), Eqs. (\ref{eq: trasf. coord.}) define the
motion of this mass with respect to the introduced frame of the local
observer. In concrete terms one gets \begin{equation}
\begin{array}{c}
x(t)=l_{1}+\frac{1}{2}l_{1}\Phi(t)+\frac{1}{2}l_{1}l_{3}\dot{\Phi}(t)\\
\\y(t)=l_{2}+\frac{1}{2}l_{2}\Phi(t)+\frac{1}{2}l_{2}l_{3}\dot{\Phi}(t)\\
\\z(t)=l_{3}-\frac{1}{4}(l_{1}^{2}+l_{2}^{2})\dot{\Phi}(t).\end{array}\label{eq: Grishuk 0}\end{equation}
 In absence of GWs the position of the mass is $(l_{1},l_{2},l_{3}).$
The effect of the scalar GW is to drive the mass to have oscillations.
Thus, in general, from Eqs. (\ref{eq: Grishuk 0}) all three components
of motion are present.

Neglecting the terms with $\dot{\Phi}$ in Eqs. (\ref{eq: Grishuk 0}),
the {}``traditional'' equations for the mass motion are obtained:\begin{equation}
\begin{array}{c}
x(t)=l_{1}+\frac{1}{2}l_{1}\Phi(t)\\
\\y(t)=l_{2}+\frac{1}{2}l_{2}\Phi(t)\\
\\z(t)=l_{3}.\end{array}\label{eq: traditional}\end{equation}
Clearly, this is the analogous of the \emph{electric} component of
motion in electrodynamics, see the Introduction of this paper and
Refs. \cite{key-38,key-39,key-40,key-41,key-42,key-43}, while equations\begin{equation}
\begin{array}{c}
x(t)=l_{1}+\frac{1}{2}l_{1}l_{3}\dot{\Phi}(t)\\
\\y(t)=l_{2}+\frac{1}{2}l_{2}l_{3}\dot{\Phi}(t)\\
\\z(t)=l_{3}-\frac{1}{4}(l_{1}^{2}+l_{2}^{2})\dot{\Phi}(t),\end{array}\label{eq: news}\end{equation}

are the analogue of the \emph{magnetic} component of motion. The fundamental
fact to be stressed is that the \emph{magnetic} component becomes
important when the frequency of the wave increases, but only in the
low-frequency regime. This can be understood directly from eqs. (\ref{eq: Grishuk 0}).
In fact, recalling that $\Phi=\Phi_{0}e^{i\omega(t+z))}$eqs. (\ref{eq: Grishuk 0})
become\begin{equation}
\begin{array}{c}
x(t)=l_{1}+\frac{1}{2}l_{1}\Phi(t)+\frac{1}{2}l_{1}l_{3}\omega\Phi(\omega t-\frac{\pi}{2})\\
\\y(t)=l_{2}+\frac{1}{2}l_{2}\Phi(t)+\frac{1}{2}l_{2}l_{3}\omega\Phi(\omega t-\frac{\pi}{2})\\
\\z(t)=l_{3}-\frac{1}{4}(l_{1}^{2}+l_{2}^{2})\omega\Phi(\omega t-\frac{\pi}{2}).\end{array}\label{eq: Grishuk 01}\end{equation}

Thus, the terms with $\dot{\Phi}$ in eqs. (\ref{eq: Grishuk 0})
can be neglected only when the wavelength goes to infinity, while,
at high-frequencies, the expansion in terms of $\omega l_{i}l_{j}$
corrections, with $i,j=1,2,3,$ breaks down.

\section{Detectability of the \emph{electric} component}

In the literature of scalar GWs, in general, the detectability is
discussed only in the low frequency-approximation, i.e. only for the
\emph{electric} component of eqs. (\ref{eq: traditional}), see \cite{key-52,key-58}
for example. 

In this case, it is well known that the geodesic deviation equation
(\ref{eq: deviazione geodetiche}) gives \cite{key-47}

\begin{equation}
\ddot{x}=\frac{1}{2}\ddot{\Phi}x\label{eq: accelerazione mareale lungo x}\end{equation}

and

\begin{equation}
\ddot{y}=\frac{1}{2}\ddot{\Phi}y.\label{eq: accelerazione mareale lungo y}\end{equation}

At this point, one can write \cite{key-59}

\begin{equation}
\widetilde{R}_{0j0}^{i}=\frac{1}{2}\left(\begin{array}{cccc}
-\partial_{t}^{2} & 0 & 0 & {}\\
0 & -\partial_{t}^{2} & 0 & \vdots\\
0 & 0 & 0 & {}\\
{} & \cdots & {} & {}\end{array}\right)\Phi(t,z)=-\frac{1}{2}T_{ij}\partial_{t}^{2}\Phi\label{eq: eqs}\end{equation}

Here the transverse projector in respect to the direction of propagation
of the GW, $\widehat{n}$, defined by \cite{key-59}

\begin{equation}
T_{ij}=\delta_{ij}-\widehat{n}_{i}\widehat{n}_{j}\;,\label{eq: Tij}\end{equation}

has been used. In this way, the geodesic deviation equation (\ref{eq: deviazione geodetiche})
can be re-written like

\begin{equation}
\frac{d^{2}}{dt^{2}}x_{i}=\frac{1}{2}\partial_{t}^{2}\Phi T_{ij}x_{j}.\label{eq: TL}\end{equation}

Concerning the detectability of the third polarization state let us
compute the pattern function of a detector to this scalar component.
One has to recall that it is possible to associate to a detector a
\textit{detector tensor} \cite{key-59} that, for an interferometer
with arms along the $\hat{u}$ e $\hat{v}$ directions with respect
the propagating gravitational wave (see Fig. 1), is defined by 

\begin{equation}
D^{ij}\equiv\frac{1}{2}(\hat{v}^{i}\hat{v}^{j}-\hat{u}^{i}\hat{u}^{j}).\label{eq: definizione D}\end{equation}

\begin{figure}
\includegraphics{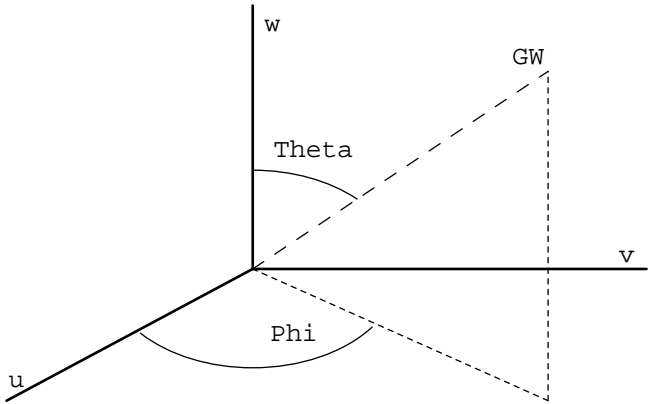}

\caption{a GW propagating from an arbitrary direction $(r,\theta,\phi)$, adapted
from Ref. \cite{key-59}}

\end{figure}
If the detector is an interferometer, the signal induced by a gravitational
wave of a generic polarization, here labelled with $s(t),$ is the
phase shift, which is proportional to \cite{key-59}

\begin{equation}
s(t)\sim D^{ij}\widetilde{R}_{i0j0}\;.\label{eq: legame onda-output}\end{equation}

Then, by using Eqs. (\ref{eq: eqs}) one gets

\begin{equation}
s(t)\sim-\sin^{2}\theta\cos2\phi\;.\label{eq: legame onda-output 2}\end{equation}

The angular dependence (\ref{eq: legame onda-output 2}), which is
shown in Fig. 2, is different from the two well-known standard ones
arising from general relativity which are, respectively $(1+\cos^{2}\theta)\cos2\phi$
for the $+$ polarization and $-\cos\theta\sin2\vartheta$ for the
$\times$ polarization, see for example Ref. \cite{key-60}. Thus,
in principle, the angular dependence (\ref{eq: legame onda-output 2})
could be used to understand if this third polarization is present,
under the expectation that the current or future GW detectors can
achieve a high sensitivity.

\begin{figure}
\includegraphics{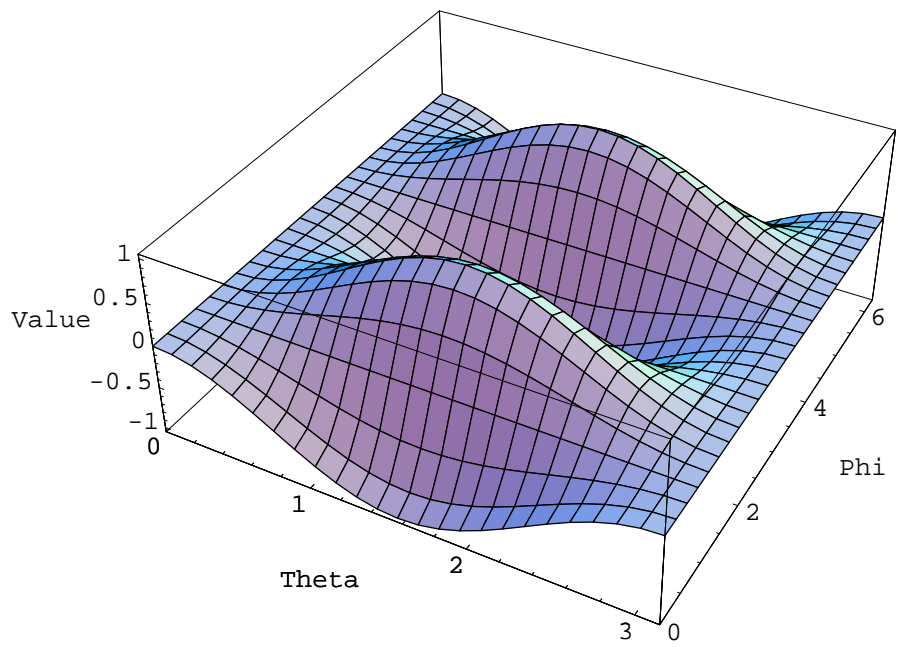}

\caption{Angular dependence of the response function for the third polarization,
adapted from Ref. \cite{key-59}}

\end{figure}

For a sake of completeness, it is better to show similar figures for
the cases of $+$ and $\times$ tensor GWs to compare with figure
2 \cite{key-20}. The angular dependences $(1+\cos^{2}\theta)\cos2\phi$
for the $+$ polarization and $-\cos\theta\sin2\vartheta$ for the
$\times$ polarization are respectively shown in figure 3 and figure
4. 

\begin{figure}
\includegraphics{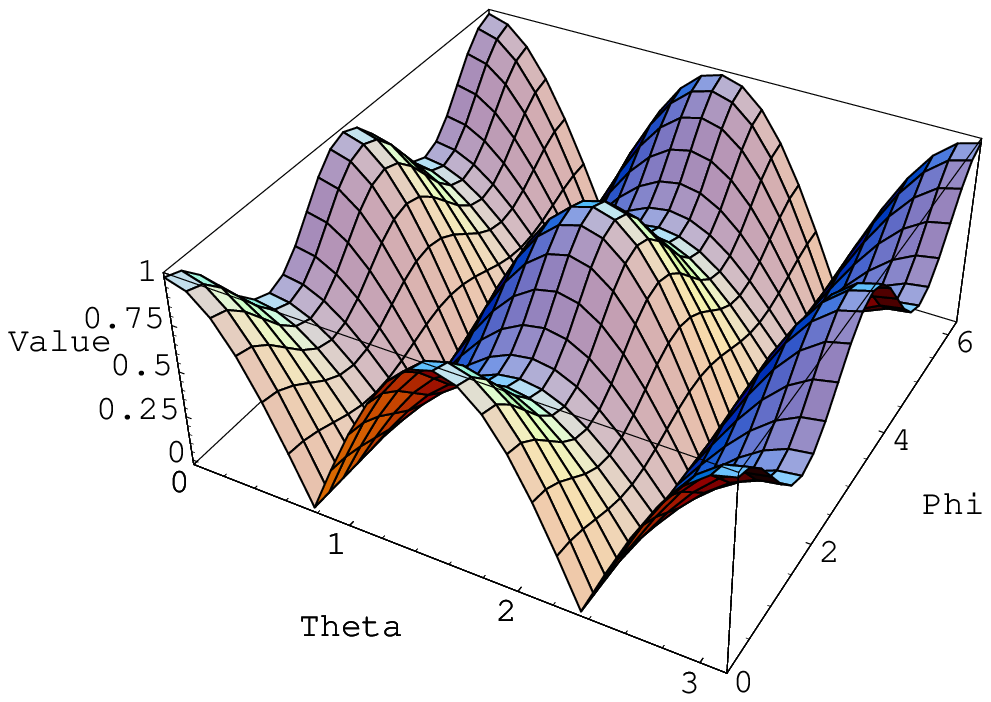}

\caption{the angular dependence $(1+\cos^{2}\theta)\cos2\phi$ for the $+$
polarization}

\end{figure}
\begin{figure}
\includegraphics{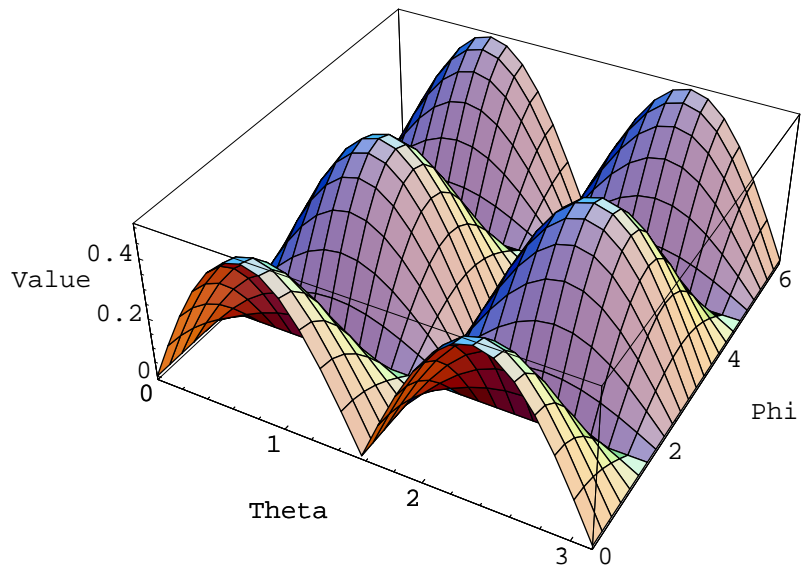}

\caption{the angular dependence $-\cos\theta\sin2\vartheta$ for the $\times$
polarization }

\end{figure}

\section{Detectability of the \emph{magnetic} component}

The discussion of previous Section concerns only the low-frequency
approximation of the \emph{electric} component of Eqs. (\ref{eq: traditional}).
For a better approximation in the response function one needs a frequency
dependence by considering the \emph{magnetic} component of Eqs. (\ref{eq: news})
too. We emphasize that in this Section 5 and in Section 6, we will
consider only the \emph{magnetic} component of scalar GWs. Notice
that we are not claiming that the electric component can be neglected
\cite{key-20}. The electric component is \emph{always} present. The
key point is that we have discusses his potential detection in Section
3. But, as we are within the linearized theory, we can invoke the
Principle of Superposition in order to discuss them separately. The
same happens when one discusses separately the various different polarizations.

To compute the response functions for an arbitrary propagating direction
of the GW a spatial rotation of the coordinate system has to be performed
\cite{key-7,key-43}:

\begin{equation}
\begin{array}{ccc}
u & = & -x\cos\theta\cos\phi+y\sin\phi+z\sin\theta\cos\phi\\
\\v & = & -x\cos\theta\sin\phi-y\cos\phi+z\sin\theta\sin\phi\\
\\w & = & x\sin\theta+z\cos\theta,\end{array}\label{eq: rotazione}\end{equation}

or, in terms of the $x,y,z$ frame:

\begin{equation}
\begin{array}{ccc}
x & = & -u\cos\theta\cos\phi-v\cos\theta\sin\phi+w\sin\theta\\
\\y & = & u\sin\phi-v\cos\phi\\
\\z & = & u\sin\theta\cos\phi+v\sin\theta\sin\phi+w\cos\theta.\end{array}\label{eq: rotazione 2}\end{equation}

The test masses are the beam splitter and the mirror of the interferometer,
and we will suppose the the beam splitter located in the origin of
the coordinate system. In this way, Eqs. (\ref{eq: news}) represent
the motion of the mirror like it is due to the \emph{magnetic} component
of the SGW.

As the mirror of Eqs. (\ref{eq: news}) is situated in the $u$ direction,
using Eqs. (\ref{eq: news}), (\ref{eq: rotazione}) and (\ref{eq: rotazione 2})
the $u$ coordinate of the mirror is given by

\begin{equation}
u=L+\frac{1}{4}L^{2}A\dot{\Phi}(t),\label{eq: du}\end{equation}

where \begin{equation}
A\equiv2\cos\theta\cos\phi[(\frac{1+\sin^{2}\theta}{2})+\sin^{2}\theta\sin2\phi]-2\sin^{2}\phi\sin\theta\cos\phi\label{eq: A-1}\end{equation}

and $L=\sqrt{l_{1}^{2}+l_{2}^{2}+l_{3}^{2}}$ is the length of the
interferometer arms.

The computation for the $v$ arm is similar to the one above. Using
Eqs. (\ref{eq: news}), (\ref{eq: rotazione}) and (\ref{eq: rotazione 2}),
the coordinate of the mirror in the $v$ arm is:

\begin{equation}
v=L+\frac{1}{4}L^{2}B\dot{\Phi}(t),\label{eq: dv}\end{equation}

where\begin{equation}
B\equiv2\cos\theta\sin\phi[(\frac{1+\sin^{2}\theta}{2})+\sin^{2}\theta\sin2\phi]-2\cos^{2}\phi\sin\theta\sin\phi.\label{eq: B}\end{equation}

Equations (\ref{eq: du}) and (\ref{eq: dv}) represent the distance
of the two mirrors of the interferometer from the beam-splitter in
presence of the scalar GW polarization (again note that only the contribution
of the \emph{magnetic} component of the third polarization of the
GW is taken into account). 

A {}``signal'' can also be defined in the time domain (i.e. $T=L$
in our notation):

\begin{equation}
\frac{\delta T(t)}{T}\equiv\frac{u-v}{L}=\frac{1}{4}L(A-B)\dot{\Phi}(t).\label{eq: signal piu}\end{equation}

The quantity (\ref{eq: signal piu}) can be computed in the frequency
domain using the Fourier transform of $\Phi$, defined by \cite{key-3}

\begin{equation}
\tilde{\Phi}(\omega)=\int_{-\infty}^{\infty}dt\Phi(t)\exp(i\omega t),\label{eq: trasformata di fourier}\end{equation}
obtaining

\[
\frac{\tilde{\delta}T(\omega)}{T}=H_{magn}^{\Phi}(\omega)\Phi(\omega),\]

where the function

\begin{equation}
\begin{array}{c}
H_{magn}^{\Phi}(\omega)=-\frac{1}{8}i\omega L(A-B)=\\
\\-\frac{1}{4}i\omega L\{\cos\theta[(\frac{1+\sin^{2}\theta}{2})+\sin^{2}\theta\sin2\phi](\cos\phi-\sin\phi)+\\
\\+\sin\theta[\cos^{2}\phi\sin\phi-\sin^{2}\phi\cos\phi]\}\end{array}\label{eq: risposta totale}\end{equation}

is the total response function of the interferometer for the \emph{magnetic}
component of the third polarization of the scalar GW. This response
function is different from the result of \cite{key-43} because we
corrected the error in Eqs. (20) of \cite{key-43} (Eqs. (\ref{eq: trasf. coord. errata})
in this paper) and we used the correct Eqs. (\ref{eq: trasf. coord.}).
Such an error was dragged along all the computations in \cite{key-43}
and this enabled incorrect geometric factors in the response function
in \cite{key-43}.

\section{A more precise response function for the \emph{magnetic} component}

Again, it is important to stress the importance of the \emph{magnetic}
component at high frequency \cite{key-20}. In fact, it is well known
that the frequency-range for earth based gravitational antennas is
the interval $10Hz\leq f\leq10KHz$ \cite{key-1}. As we recalled
in the introduction, the \emph{magnetic} contribution represents the
finite-wavelength correction to the usual infinite-wavelength approximation.
In other words, it becomes important at high frequencies, i.e, frequencies
at order KHzs \cite{key-38,key-39,key-40,key-41,key-42,key-43}. Thus,
in this Section a more precise response function for the \emph{magnetic}
component at high frequency will be obtained. 

Following \cite{key-3,key-19,key-60,key-61,key-62}, a good way to
analyse variations in the proper distance (time) is by means of {}``bouncing
photons''. A photon can be launched from the interferometer's beam-splitter
to be bounced back by the mirror. The {}``bouncing photons analysis''
was created in \cite{key-61}. Actually, it has strongly generalized
to angular dependences and scalar waves in \cite{key-3,key-19,key-60,key-62}.
However, this is the first time that the such a {}``bouncing photons
analysis'' is applied to the \emph{magnetic} component of scalar
GWs.

We start by considering a photon which propagates in the $u$ axis,
but the analysis is almost the same for a photon which propagates
in the $v$ axis. By using eq. (\ref{eq: du}), the unperturbed coordinates
for the beam-splitter and the mirror are $u_{b}=0$ and $u_{m}=L$.
Thus, the unperturbed propagation time between the two masses is

\begin{equation}
T=L.\label{eq: tempo imperturbato}\end{equation}

From eq. (\ref{eq: du}), the displacements of the two masses under
the influence of the GW are

\begin{equation}
\delta u_{b}(t)=0\label{eq: spostamento beam-splitter}\end{equation}

and

\begin{equation}
\delta u_{m}(t)=\frac{1}{4}L^{2}A\dot{\Phi}(t+L\sin\theta\cos\phi).\label{eq: spostamento mirror}\end{equation}

In this way, the relative displacement in the $u$ direction, which
is defined by

\begin{equation}
\delta L(t)=\delta u_{m}(t)-\delta u_{b}(t)\label{eq: spostamento relativo}\end{equation}

gives a {}``signal'' in the $u$ direction

\begin{equation}
\frac{\delta T(t)}{T}|_{u}=\frac{\delta L(t)}{L}=\frac{1}{4}LA\dot{\Phi}(t+L\sin\theta\cos\phi).\label{eq: strain magnetico}\end{equation}
But, for a large separation between the test masses (in the case of
Virgo the distance between the beam-splitter and the mirror is three
kilometres, four in the case of LIGO), the definition (\ref{eq: spostamento relativo})
for relative displacements becomes unphysical because the two test
masses are taken at the same time and therefore cannot be in a casual
connection \cite{key-61,key-62}. In this way, the correct definitions
for the bouncing photon are

\begin{equation}
\delta L_{1}(t)=\delta u_{m}(t)-\delta u_{b}(t-T_{1})\label{eq: corretto spostamento B.S. e M.}\end{equation}

and

\begin{equation}
\delta L_{2}(t)=\delta u_{m}(t-T_{2})-\delta u_{b}(t),\label{eq: corretto spostamento B.S. e M. 2}\end{equation}

where $T_{1}$ and $T_{2}$ are the photon propagation times for the
forward and return trip correspondingly. According to the new definitions,
the displacement of one test mass is compared with the displacement
of the other at a later time to allow for finite delay from the light
propagation. The propagation times $T_{1}$ and $T_{2}$ in Eqs. (\ref{eq: corretto spostamento B.S. e M.})
and (\ref{eq: corretto spostamento B.S. e M. 2}) can be replaced
with the nominal value $T$ because the test mass displacements are
already first order in $\dot{\Phi}$ \cite{key-62}. Thus, the total
change in the distance between the beam splitter and the mirror in
one round-trip of the photon is

\begin{equation}
\delta L_{r.t.}(t)=\delta L_{1}(t-T)+\delta L_{2}(t)=2\delta u_{m}(t-T)-\delta u_{b}(t)-\delta u_{b}(t-2T),\label{eq: variazione distanza propria}\end{equation}

and in terms of the amplitude of the scalar GW:

\begin{equation}
\delta L_{r.t.}(t)=\frac{1}{2}L^{2}A\dot{\Phi}(t+L\sin\theta\cos\phi-L).\label{eq: variazione distanza propria 2}\end{equation}
The change in distance (\ref{eq: variazione distanza propria 2})
leads to changes in the round-trip time for photons propagating between
the beam-splitter and the mirror in the $u$ direction:

\begin{equation}
\frac{\delta_{1}T(t)}{T}|_{u}=\frac{1}{2}LA\dot{\Phi}(t+L\sin\theta\cos\phi-L).\label{eq: variazione tempo proprio 1}\end{equation}

In the last calculation (variations in the photon round-trip time
which come from the motion of the test masses inducted by the magnetic
component of the scalar GW), it has been implicitly assumed that the
propagation of the photon between the beam-splitter and the mirror
of our interferometer is uniform as if it were moving in a flat space-time.
But the presence of the tidal forces indicates that the space-time
is curved. As a result, one more effect after the first discussed,
that requires spacial separation, has to be analysed \cite{key-61,key-62}. 

From equation (\ref{eq: spostamento mirror}) the tidal acceleration
of a test mass caused by the magnetic component of the $+$ polarization
of the GW in the $u$ direction is \begin{equation}
\ddot{u}(t+u\sin\theta\cos\phi)=\frac{1}{4}L^{2}A\frac{\partial}{\partial t}\ddot{\Phi}(t+u\sin\theta\cos\phi).\label{eq: acc}\end{equation}

Equivalently, one can say that there is a gravitational potential
\cite{key-22,key-61,key-62}:

\begin{equation}
V(u,t)=-\frac{1}{4}L^{2}A\int_{0}^{u}\frac{\partial}{\partial t}\ddot{\Phi}(t+l\sin\theta\cos\phi)dl,\label{eq:potenziale in gauge Lorentziana}\end{equation}

which generates the tidal forces, and that the motion of the test
mass is governed by the Newtonian equation \cite{key-22,key-61,key-62}

\begin{equation}
\ddot{\overrightarrow{r}}=-\bigtriangledown V.\label{eq: Newtoniana}\end{equation}

For the second effect one considers the interval for photons propagating
along the $u$ -axis\begin{equation}
ds^{2}=g_{00}dt^{2}+du^{2}.\label{eq: metrica osservatore locale}\end{equation}

The condition for a null trajectory ($ds=0$) gives the coordinate
velocity of the photons \cite{key-61,key-62}

\begin{equation}
v_{p}^{2}\equiv(\frac{du}{dt})^{2}=1+2V(t,u),\label{eq: velocita' fotone in gauge locale}\end{equation}

which to first order in $\Phi$is approximated by

\begin{equation}
v_{p}\approx\pm[1+V(t,u)],\label{eq: velocita fotone in gauge locale 2}\end{equation}

with $+$ and $-$ for the forward and return trip respectively. By
knowing the coordinate velocity of the photon, one defines the propagation
time for its travelling between the beam-splitter and the mirror:

\begin{equation}
T_{1}(t)=\int_{u_{b}(t-T_{1})}^{u_{m}(t)}\frac{du}{v_{p}}\label{eq:  tempo di propagazione andata gauge locale}\end{equation}

and

\begin{equation}
T_{2}(t)=\int_{u_{m}(t-T_{2})}^{u_{b}(t)}\frac{(-du)}{v_{p}}.\label{eq:  tempo di propagazione ritorno gauge locale}\end{equation}

The calculations of these integrals would be complicated because the
$u_{m}$ boundaries of them are changing with time:

\begin{equation}
u_{b}(t)=0\label{eq: variazione b.s. in gauge locale}\end{equation}

and

\begin{equation}
u_{m}(t)=L+\delta u_{m}(t).\label{eq: variazione specchio nin gauge locale}\end{equation}

But, to first order in $\Phi,$ these contributions can be approximated
by $\delta L_{1}(t)$ and $\delta L_{2}(t)$ (see Eqs. (\ref{eq: corretto spostamento B.S. e M.})
and (\ref{eq: corretto spostamento B.S. e M. 2})). Thus, the combined
effect of the varying boundaries is given by $\delta_{1}T(t)$ in
eq. (\ref{eq: variazione tempo proprio 1}). Then, only the times
for photon propagation between the fixed boundaries, i.e $0$ and
$L$, have to be calculated. Such propagation times are denoted with
$\Delta T_{1,2}$ to distinguish from $T_{1,2}$. In the forward trip,
the propagation time between the fixed limits is

\begin{equation}
\Delta T_{1}(t)=\int_{0}^{L}\frac{du}{v(t',u)}\approx L-\int_{0}^{L}V(t',u)du,\label{eq:  tempo di propagazione andata  in gauge locale}\end{equation}

where $t'$ is the delay time (i.e. $t$ is the time at which the
photon arrives in the position $L$, so $L-u=t-t'$) which corresponds
to the unperturbed photon trajectory: 

\begin{center}
$t'=t-(L-u)$. 
\par\end{center}

Similarly, the propagation time in the return trip is

\begin{equation}
\Delta T_{2}(t)=L-\int_{L}^{0}V(t',u)du,\label{eq:  tempo di propagazione andata  in gauge locale 2}\end{equation}

where now the delay time is given by

\begin{center}
$t'=t-u$.
\par\end{center}

The sum of $\Delta T_{1}(t-T)$ and $\Delta T_{2}(t)$ gives the round-trip
time for photons travelling between the fixed boundaries. Then, the
deviation of this round-trip time (distance) from its unperturbed
value $2T$ is\begin{equation}
\begin{array}{c}
\delta_{2}T(t)=-\int_{0}^{L}[V(t-2L+u,u)du+\\
\\-\int_{L}^{0}V(t-u,u)]du,\end{array}\label{eq: variazione tempo proprio 2}\end{equation}

and, using Eq. (\ref{eq:potenziale in gauge Lorentziana}), it is

\begin{equation}
\begin{array}{c}
\delta_{2}T(t)=\frac{1}{4}L^{2}A\int_{0}^{L}[\int_{0}^{u}\frac{\partial}{\partial t}\ddot{\Phi}(t-2T+l(1+\sin\theta\cos\phi))dl+\\
\\-\int_{0}^{u}\frac{\partial}{\partial t}\ddot{\Phi}(t-l(1-\sin\theta\cos\phi)dl]du.\end{array}\label{eq: variazione tempo proprio 2 rispetto h}\end{equation}

Thus, the total round-trip proper distance in presence of the \emph{magnetic}
component of the scalar GW is:

\begin{equation}
T_{t}=2T+\delta_{1}T+\delta_{2}T,\label{eq: round-trip  totale in gauge locale}\end{equation}

and\begin{equation}
\delta T_{u}=T_{t}-2T=\delta_{1}T+\delta_{2}T\label{eq:variaz round-trip totale in gauge locale}\end{equation}

is the total variation of the proper time (distance) for the round-trip
of the photon in presence of the \emph{magnetic} component of the
scalar GW in the $u$ direction.

By using Eqs. (\ref{eq: variazione tempo proprio 1}), (\ref{eq: variazione tempo proprio 2 rispetto h})
and the Fourier transform of $\Phi$ defined by Eq. (\ref{eq: trasformata di fourier}),
the quantity (\ref{eq:variaz round-trip totale in gauge locale})
can be computed in the frequency domain as 

\begin{equation}
\tilde{\delta}T_{u}(\omega)=\tilde{\delta}_{1}T(\omega)+\tilde{\delta}_{2}T(\omega)\label{eq:variaz round-trip totale in gauge locale 2}\end{equation}

where

\begin{equation}
\tilde{\delta}_{1}T(\omega)=-i\omega\exp[i\omega L(1-\sin\theta\cos\phi)]\frac{L^{2}A}{2}\tilde{\Phi}(\omega)\label{eq: dt 1 omega}\end{equation}

\begin{equation}
\begin{array}{c}
\tilde{\delta}_{2}T(\omega)=\frac{i\omega L^{2}A}{4}[\frac{-1+\exp[i\omega L(1-\sin\theta\cos\phi)]-iL\omega(1-\sin\theta\cos\phi)}{(1-\sin\theta\cos\phi)^{2}}+\\
\\+\frac{\exp(2i\omega L)(1-\exp[i\omega L(-1-\sin\theta\cos\phi)]-iL\omega(1+\sin\theta\cos\phi)}{(-1-\sin\theta\cos\phi)^{2}}]\tilde{\Phi}(\omega).\end{array}\label{eq: dt 2 omega}\end{equation}

In the above computation the derivation and translation theorems of
the Fourier transform have been used. In this way the response function
of the $u$ arm of our interferometer to the magnetic component of
the scalar GW results

\begin{equation}
\begin{array}{c}
H_{u}^{\Phi}(\omega)\equiv\frac{\tilde{\delta}T_{u}(\omega)}{L\tilde{\Phi}(\omega)}=\\
\\=-i\omega\exp[i\omega L(1-\sin\theta\cos\phi)]\frac{LA}{2}+\\
\\\frac{i\omega LA}{4}[\frac{-1+\exp[i\omega L(1-\sin\theta\cos\phi)]-iL\omega(1-\sin\theta\cos\phi)}{(1-\sin\theta\cos\phi)^{2}}+\\
\\+\frac{\exp(2i\omega L)(1-\exp[i\omega L(-1-\sin\theta\cos\phi)]-iL\omega(1+\sin\theta\cos\phi)}{(-1-\sin\theta\cos\phi)^{2}}].\end{array}\label{eq: risposta u}\end{equation}

The computation for the $v$ arm is parallel to the one above. With
the same way of thinking of previous analysis, one gets variations
in the photon round-trip time which come from the motion of the beam-splitter
and the mirror in the $v$ direction:

\begin{equation}
\frac{\delta_{1}T(t)}{T}|_{v}=\frac{1}{2}LB\Phi(t+L\sin\theta\sin\phi-L),\label{eq: variazione tempo proprio 1 in v}\end{equation}

while the second contribute (propagation in a curve spacetime) will
be 

\begin{equation}
\begin{array}{c}
\delta_{2}T(t)=\frac{1}{4}L^{2}B\int_{0}^{L}[\int_{0}^{u}\frac{\partial}{\partial t}\ddot{\Phi}(t-2T+l(1-\sin\theta\sin\phi))dl+\\
\\-\int_{0}^{u}\frac{\partial}{\partial t}\ddot{\Phi}(t-l(1-\sin\theta\sin\phi)dl]du,\end{array}\label{eq: variazione tempo proprio 2 rispetto h in v}\end{equation}

and the total response function of the $v$ arm for the \emph{magnetic}
component of the scalar GWs is given by\begin{equation}
\begin{array}{c}
H_{v}^{\Phi}(\omega)\equiv\frac{\tilde{\delta}T_{u}(\omega)}{L\ddot{\Phi}\omega)}=\\
\\=-i\omega\exp[i\omega L(1-\sin\theta\sin\phi)]\frac{LB}{2}+\\
\\+\frac{i\omega LB}{4}[\frac{-1+\exp[i\omega L(1-\sin\theta\sin\phi)]-iL\omega(1-\sin\theta\sin\phi)}{(1-\sin\theta\cos\phi)^{2}}+\\
\\+\frac{\exp(2i\omega L)(1-\exp[i\omega L(-1-\sin\theta\sin\phi)]-iL\omega(1+\sin\theta\sin\phi)}{(-1-\sin\theta\sin\phi)^{2}}].\end{array}\label{eq: risposta v}\end{equation}

The total response function for the \emph{magnetic} component is given
by the difference of the two response function of the two arms:\begin{equation}
H_{tot}^{\Phi}(\omega)\equiv H_{u}^{\Phi}(\omega)-H_{v}^{\Phi}(\omega),\label{eq: risposta totalissima}\end{equation}

and using Eqs. (\ref{eq: risposta u}) and (\ref{eq: risposta v})
one obtains a complicated formula\begin{equation}
\begin{array}{c}
H_{tot}^{\Phi}(\omega)=\frac{\tilde{\delta}T_{tot}(\omega)}{L\tilde{\Phi}(\omega)}=\\
\\=-i\omega\exp[i\omega L(1-\sin\theta\cos\phi)]\frac{LA}{2}+\frac{LB}{2}i\omega\exp[i\omega L(1-\sin\theta\sin\phi)]\\
\\-\frac{i\omega LA}{4}[\frac{-1+\exp[i\omega L(1-\sin\theta\cos\phi)]-iL\omega(1-\sin\theta\cos\phi)}{(1-\sin\theta\cos\phi)^{2}}\\
\\+\frac{\exp(2i\omega L)(1-\exp[i\omega L(-1-\sin\theta\cos\phi)]-iL\omega(1+\sin\theta\cos\phi)}{(-1-\sin\theta\cos\phi)^{2}}]+\\
\\+\frac{i\omega LB}{4}[\frac{-1+\exp[i\omega L(1-\sin\theta\sin\phi)]-iL\omega(1-\sin\theta\sin\phi)}{(1-\sin\theta\cos\phi)^{2}}+\\
\\+\frac{\exp(2i\omega L)(1-\exp[i\omega L(-1-\sin\theta\sin\phi)]-iL\omega(1+\sin\theta\sin\phi)}{(-1-\sin\theta\sin\phi)^{2}}],\end{array}\label{eq: risposta totale 2}\end{equation}

that, at lower frequencies is in perfect agreement with the result
(\ref{eq: risposta totale}):

\begin{equation}
\begin{array}{c}
H_{tot}^{\Phi}(\omega\rightarrow0)=\\
\\=-\frac{1}{4}i\omega L\{\cos\theta[(\frac{1+\sin^{2}\theta}{2})+\sin^{2}\theta\sin2\phi](\cos\phi-\sin\phi)+\\
\\+\sin\theta[\cos^{2}\phi\sin\phi-\sin^{2}\phi\cos\phi]\}\end{array}\label{eq: risposta totale bassa}\end{equation}

In figure 5 the angular dependence (\ref{eq: risposta totale 2})
is mapped at a frequency of $9KHz$ for the Virgo interferometer ($L=3km$,
see \cite{key-1}). From figure 5 it is clear why we are claiming
that the \emph{magnetic} contribution becomes important at high frequencies:
if one neglects such a contribution considering only the low-frequency
approximation of the \emph{electric} contribution analysed in previous
literature and in Section 4 of this paper an important portion of
the total integrated signal could be, in principle, lost. In fact,
the lost signal could arrive at about the $15\%$ for some particular
directions of the propagating GW. To well understand this point one
has to compare this \emph{magnetic} contribution, which is shown in
figure 5, with the \emph{electric} contribution which is shown in
figure 2, that is sufficient only for frequencies order hundreds Hz.
For higher frequencies, i.e. frequencies order kHzs, the \emph{magnetic}
correction is needed.

\begin{figure}

\includegraphics{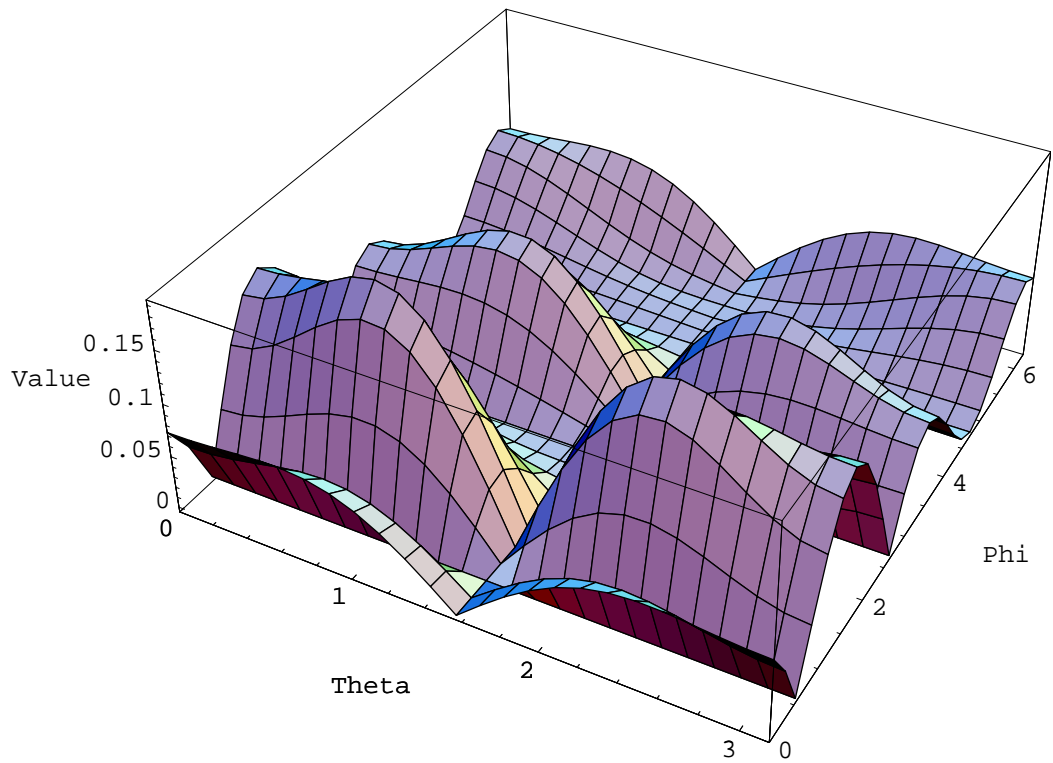}\caption{The angular dependence of the \emph{magnetic} response function (\ref{eq: risposta totale 2})
at 9KHz for the Virgo interferometer ($L=3km$) }

\end{figure}

\section{Comparing with General Theory of Relativity}

It is important to show an expansion of the main results recalling
its presence also in GRT \cite{key-44}. Doing that, the importance
of the STG for the effect, that is known to exist also in GRT, is
further emphasized \cite{key-44}. To make this, let us insert in
Eqs. (\ref{eq: Grishuk 0}) the contribution due to the $+$ and $\times$
polarizations of the the total perturbation (\ref{eq: perturbazione totale}).
The analogous of Eqs. (\ref{eq: Grishuk 0}) for the $+$ and $\times$
polarizations in GTR are Eqs. (6) of Ref. \cite{key-39}, which are 

\begin{equation}
\begin{array}{c}
x(t)=l_{1}+\frac{1}{2}[l_{1}h_{+}(t)-l_{2}h_{\times}(t)]+\frac{1}{2}l_{1}l_{3}\dot{h}_{+}(t)+\frac{1}{2}l_{2}l_{3}\dot{h}_{\times}(t)\\
\\y(t)=l_{2}-\frac{1}{2}[l_{2}h_{+}(t)+l_{1}h_{\times}(t)]-\frac{1}{2}l_{2}l_{3}\dot{h}_{+}(t)+\frac{1}{2}l_{1}l_{3}\dot{h}_{\times}(t)\\
\\z(t)=l_{3}-\frac{1}{4}(l_{1}^{2}-l_{2}^{2})\dot{h}_{+}(t)+2l_{1}l_{2}\dot{h}_{\times}(t).\end{array}\label{eq: Grishuk GTR}\end{equation}

These equations, which are also Eqs. (13) of Ref. \cite{key-38} written
with different notations, define the motion of the mass due to the
$+$ and $\times$ polarizations in the same frame of the local observer
of Eqs. (\ref{eq: Grishuk 0}). 

Neglecting the terms with $\dot{h}_{+}$ and $\dot{h}_{\times}$ in
eqs. (\ref{eq: Grishuk GTR}), the {}``\emph{traditional}'' equations
for the mass motion in GTR are obtained \cite{key-38,key-39}:\begin{equation}
\begin{array}{c}
x(t)=l_{1}+\frac{1}{2}[l_{1}h_{+}(t)-l_{2}h_{\times}(t)]\\
\\y(t)=l_{2}-\frac{1}{2}[l_{2}h_{+}(t)+l_{1}h_{\times}(t)]\\
\\z(t)=l_{3}.\end{array}\label{eq: traditional GTR}\end{equation}

Clearly, this is the analogous of the electric component of motion
in electrodynamics \cite{key-38,key-39}, while equations\begin{equation}
\begin{array}{c}
x(t)=l_{1}+\frac{1}{2}l_{1}l_{3}\dot{h}_{+}(t)+\frac{1}{2}l_{2}l_{3}\dot{h}_{\times}(t)\\
\\y(t)=l_{2}-\frac{1}{2}l_{2}l_{3}\dot{h}_{+}(t)+\frac{1}{2}l_{1}l_{3}\dot{h}_{\times}(t)\\
\\z(t)=l_{3}-\frac{1}{4}(l_{1}^{2}-l_{2}^{2})\dot{h}_{+}(t)+2l_{1}l_{2}\dot{h}_{\times}(t),\end{array}\label{eq: untraditional GTR}\end{equation}

are the analogue of the magnetic component of motion \cite{key-38,key-39}.
Starting from Eqs. (\ref{eq: untraditional GTR}), a careful analysis
has been realized in \cite{key-39} where the response functions for
the \emph{magnetic} components in GTR have been computed \cite{key-20}.
In particular, the analogous of Eq. (\ref{eq: risposta totale}) for
the $+$ and $\times$ polarizations are respectively \cite{key-39}

\begin{equation}
\begin{array}{c}
H_{magn}^{+}(\omega)=-\frac{1}{8}i\omega L(A-B)=\\
\\=-\frac{1}{4}i\omega L\sin\theta[(\cos^{2}\theta+\sin2\phi\frac{1+\cos^{2}\theta}{2})](\cos\phi-\sin\phi)\end{array}\label{eq: risposta totale GTR}\end{equation}

and \begin{equation}
\begin{array}{c}
H_{magn}^{\times}(\omega)=-i\omega T(C-D)=\\
\\=-i\omega L\sin2\phi(\cos\phi+\sin\phi)\cos\theta.\end{array}\label{eq: risposta totale 2 per}\end{equation}

By invoking the Principle of Superposition, we can add the motion
of the mass due to the third scalar polarization $\Phi$, which is
defined by Eqs. (\ref{eq: Grishuk 0}), to the motion of the to mass
due to the $+$ and $\times$ polarizations, which is defined by Eqs.
(\ref{eq: Grishuk GTR}). At the end we get

\begin{equation}
\begin{array}{c}
x(t)=l_{1}+\frac{1}{2}[l_{1}h_{+}(t)-l_{2}h_{\times}(t)]+\frac{1}{2}l_{1}l_{3}\dot{h}_{+}(t)+\frac{1}{2}l_{2}l_{3}\dot{h}_{\times}(t)+\frac{1}{2}l_{1}\Phi(t)+\frac{1}{2}l_{1}l_{3}\dot{\Phi}(t)\\
\\y(t)=l_{2}-\frac{1}{2}[l_{2}h_{+}(t)+l_{1}h_{\times}(t)]-\frac{1}{2}l_{2}l_{3}\dot{h}_{+}(t)+\frac{1}{2}l_{1}l_{3}\dot{h}_{\times}(t)+\frac{1}{2}l_{2}\Phi(t)+\frac{1}{2}l_{2}l_{3}\dot{\Phi}(t)\\
\\z(t)=l_{3}-\frac{1}{4}(l_{1}^{2}-l_{2}^{2})\dot{h}_{+}(t)+2l_{1}l_{2}\dot{h}_{\times}(t)-\frac{1}{4}(l_{1}^{2}+l_{2}^{2})\dot{\Phi}(t).\end{array}\label{eq: Corda}\end{equation}

These equations define the motion of the mass due to \emph{all} the
three $+$, $\times$ and $\Phi$ polarizations of GWs in STG.

Thus, one can interpret the linearized scalar field $\Phi$ like a
small quantity that measures the scalar sector in STG, so that when
the expansion parameter vanishes one goes over to GTR \cite{key-44}.

\section{Conclusions}

In the framework of the potential detection of GWs, the important
issue of the\emph{ magnetic} component of GWs has been considered
in various paper in the literature. The analyses on this issue have
shown that such a \emph{magnetic} component results particularly important
in the high frequency portion of the frequency range of ground based
interferometers for GWs which arises from standard GTR. On the other
hand, detectors for GWs will be important also because the interferometric
GWs detection will be the definitive test for GTR or, alternatively,
a strong endorsement for ETG. In fact, recently, the \emph{magnetic}
component has been extended to GWs arising from STG, which is an alternative
candidate to GTR. After a review of some important issues on GWs in
STG, in this paper the \emph{magnetic} component has been re-analysed
in from a different point of view, by correcting an error in a previous
paper and by releasing a more precise response function. In this way,
we have also shown that if one neglects the \emph{magnetic} contribution
considering only the low-frequency approximation of the \emph{electric}
contribution, an important portion of the signal could be, in principle,
lost. In fact, the lost signal could arrive at about the $15\%$ for
some particular directions of the propagating GW as it is clear by
comparing the total \emph{magnetic} contribution, which is shown in
figure 5, with the \emph{electric} contribution which is shown in
figure 2.

At the end of the paper an expansion of the main results has been
also shown. This point is important in order to emphasize the presence
of the magnetic component in GRT too.

\section{Acknowledgements}

The Institute for Theoretical Physics and Mathematics Einstein-Galilei
and the R. M. Santilli Foundations have to be thanked for supporting
this research. I thank Herman Mosquera Cuesta, Jeremy Dunning-Davies
and Mariafelicia De Laurentis for the useful discussions. Finally,
I have to thank a lot the referees A and B for useful comments and
advices and the referee D for his enlightening correction to the error
in Eqs. (\ref{eq: trasf. coord. errata}). The excellent work of these
referees permitted to strongly improve this paper.

\end{document}